\documentclass{aa}

\newcommand{\kepler}{{\em Kepler\/}}
\newcommand{\tess}{{\em TESS\/}}
\newcommand{\dsct}{\mbox{$\delta$~Sct}}
\newcommand{\corot}{{\em CoRoT\/}}

\usepackage{graphicx,subfigure}
\usepackage{xspace}
\usepackage{array}
\usepackage{bookmark}
\usepackage{booktabs}
\usepackage{appendix}
\bibpunct{(}{)}{;}{a}{}{,}

\usepackage{txfonts}

\usepackage{hyperref}

\begin{document}

   \title{A complex network approach to \tess\ light curves of \dsct\ stars}

   \author{E. Ziaali
          \inst{1}\fnmsep\thanks{ziaali@iaa.csic.es},
          S. de Franciscis\inst{1},
          J. Pascual-Granado\inst{1}\fnmsep\thanks{javier@iaa.csic.es},
          N. Alipour\inst{2},
          H. Safari\inst{3},
          R. Garrido\inst{1},    
          J. Rodríguez-Gómez\inst{1}}

   \institute{Instituto de Astrofísica de Andalucía (CSIC), Glorieta de la Astronomía s/n, E-18008 Granada, Spain\\
              \email{ziaali@iaa.csic.es}
         \and
             Department of Physics, University of Guilan, 41335–1914, Rasht, Iran
         \and             
            Department of Physics, Faculty of Science, University of Zanjan, 45371-38791, Zanjan, Iran         
             }

   \date{}

  \abstract 
   {"Complex systems" have many highly interconnected dynamical units that exhibit non-linear properties. \dsct\ stars, as a type of pulsating stars, have intrinsic brightness variations that need non-linear models to be described completely. These stars span a broad range of properties from low amplitude to high amplitude \dsct\ stars as well as a broad range of complex features. We applied the complex network approach to \dsct\ stars as non-linear complex systems.}
   {The differences between the constructed networks, which might appear in network metrics of low amplitude and high amplitude \dsct\ stars, can show us intrinsic asteroseismic differences that are essential for classification of pulsating stars. Additionally, the relations between the asteroseismic parameters and network metrics (such as degree or clustering distributions) can lead us to a better understanding of pulsating stars dynamics.}
   {By using the horizontal visibility algorithm, the \tess\ light curves of 69 \dsct\ stars are mapped to undirected horizontal visibility graphs (HVGs), where the graph nodes represent the light curve points. Then, the morphological characteristics of HVGs, such as the distribution of links between nodes (degree distribution) or the average fraction of triangles around each node (average clustering coefficient) are measured.}
   {The average clustering coefficients for HADS and LADS show two different linear correlations with the peak-to-peak amplitude of the \tess\ light curves that is naturally separating them into two groups, thus, avoiding an ad-hoc criterion for the first time. Exponential fits on HVG degree distributions for both HADS and LADS stars give indices suggesting correlated stochastic generating processes for \dsct\ light curves. By applying the theoretical expression for the HVGs degree distribution of random time series we can distinguish significant pulsations from the background noise which might become a practical tool in frequency analysis of stars.}
   {}

   \keywords{Asteroseismology -- Stars: oscillations -- Stars: variables: \dsct\ stars
   }
   \titlerunning{Complexity in the \dsct\ light curves}
   \authorrunning{Ziaali et al.}
   \maketitle

\section{Introduction}

Space telescopes such as \corot\ \emph{(Convection, Rotation and planetary Transits)} \citep{Baglin2009}, \kepler\ \citep{Gilliland2010} and \tess\ \emph{(Transiting Exoplanet Survey Satellite)} \citep{Ricker2015} provide the most accurate light curves for seismic and exoplanets investigations. The ultra-high precision of these satellites reveals an unprecedented detail in the sinusoidal distortions of Cepheid \citep{Christy1964}, RR Lyrae \citep{Christy1966ApJ}, \dsct\ \citep{Stellingwerf1980}, SPB and $\gamma$ Dor stars \citep{Kurtz2015MNRAS}. These are caused by non-linear mechanisms within pulsators which produce asymmetric light curves, typically, with sharp ascents and slow descents of luminosity. Combination frequencies and harmonics can also appear in the frequency domain as a consequence of the non-linear response of the stellar envelope to the excited oscillations \citep{Mariel2022}. Amplitude modulation \citep{bowman2014}, mode coupling \citep{Barcelo_Forteza2015}, tidally excited perturbations, binarity, and planets effects \citep{Steindl2021}, and non-linear driving mechanisms of modes are complex features of stars light curves. In \dsct\ stars and, in particular, in the high-amplitude \dsct\ (called HADS) subclass, the nonlinear effects in pulsations are very significant. 
  
\dsct\ stars are pulsating stars that have spectral types A0$\text{-}$F5 and effective temperatures between 6500\, K and 9500\, K. They pulsate in low-order pressure modes and they have dominant pulsation frequencies in the range of 5-80\,d$^{-1}$. \dsct\ stars have masses in a range of 1.2 and 2.5 solar masses and are an ideal class of pulsators for the
study of different phenomena in stellar physics, since they are placed in a transition region between the fastest rotators\footnote{Note, however, that cool stars are typically slower rotators than hot stars because their thick convective envelope undergoes angular momentum loss during their evolution (see e.g. \citep{Saders13}). So while rotation periods for hot stars can be less of one day, for cool stars it can become several weeks or months.} among low-mass solar-like stars, with thick convective envelopes, and the slowest rotators among high-mass stars, with large convective cores and radiative envelopes \citep{Breger2000Balt, aerts2010asteroseismology}. 

The HADS stars are classified in population $\it{I}$ of \dsct\ stars with V band amplitude $\geq 0.3$ mag and also $v\sin i \leq\ 30\ km s^{-1}$ \citep{Breger2000}. In addition, \cite{Suarez02} found a clear correlation between $v\sin i$ and oscillation amplitudes. The frequency spectrum of HADS stars mostly shows only one, two or three independent modes, which are probably radial. This can be used to characterise these stars through period ratios vs period diagrams \citep{Suarez06}. On the other hand low-amplitude \dsct\ stars (called LADS) pulsate in several modes with V band amplitudes $\leq$ 0.1 mag that could be a result of their greater rotational velocities ($v\sin i \simeq 306\ km s^{-1}$) \citep{Breger2007, Chang2013,Tim2020} (see appendix \ref{A}). In \dsct\ stars, the well-known $\kappa$ mechanism, which operates in zones of partial ionization of hydrogen and helium, can drive low-order radial and nonradial modes of the low spherical degree to measurable amplitudes (opacity-driven unstable modes). In lower temperature \dsct\ stars, near the red edge of the instability strip with substantial outer convection zones, the selection mechanism of modes with observable amplitudes could be affected by induced fluctuations of the turbulent convection. Some studies suggested the opacity-driven unstable p modes, in which non-linear processes limit their amplitudes, and intrinsically stable stochastically driven (solar-like) p modes can be excited simultaneously in the same \dsct\ star \citep{Houdek1999, Samadi2002, Antoci2011, Antoci2014}. Due to small-amplitude nonradial modes in \dsct\ light curves, categorizing the HADS and LADS stars is challenging.

A complex system is composed of a large number of highly interconnected individual dynamical units that exhibit emergent collective characteristics. The complex network approach could be used to classify features with the same characteristics (for example the degree distribution, average clustering coefficient, and shortest path length) and quantify the complexity of dynamic systems \citep{newman2003, boccaletti2006, newman2010, barabasi2016}. 

Complex network analysis have been widely applied for distinguishing individual and collective features in many different disciplines, ranging from non-linear science to biology \citep{barabasi2004network}, economics \citep{Souma2003PhyA}, engineering \citep{Ganesh2020}, earthquakes \citep{Baiesi-PRE, Pasten2018Chaos,vogel2020measuring}, Solar physics \citep{Daei2017ApJ, Gheibi2017ApJ, Lotfi2020Chaos, Mohammadi2021JGRA}, and stellar light curves \citep{munoz2021}. Graph theory as a powerful mathematical tool is used for creating the complex networks. In this work we exploit the visibility graph method in order to transform light curves into complex networks \citep{lacasa2008}. By mapping the time series into a horizontal visibility graph \citep{luque2009horizontal}, we can capture the essential characteristics of the features into distinct individual categories.
   
Here, we study the characteristics of HADS and LADS stars by mapping the light curves into the graphs (or networks) using the horizontal visibility algorithm. We study the networks local and global characteristics (nodes degree distributions, clustering coefficients, path length, etc.) for HADS and LADS stars. 
Section \ref{data} explains the sample of stars that we have used from public \tess\ data sets. Section \ref{methods} discusses the HVG algorithm. In section \ref{results} we show the results of the complex network analysis and discuss the insight they can provide in \dsct\ stars studies. Section \ref{concs} remarks on the important findings of this study.

\section{Data}\label{data}
\tess\ is the high-precision photometric instrument by NASA that is intended to hunt the Neptune- or sub-Neptune-size planets. It scans the sky in sectors with a size of  24° x 96°, where each sector lasts for two orbits of the satellite around the Earth and Moon or about 27 days on average. \tess\ observes hundreds of thousands of stars at 600-1000 nm bands in a short cadence of 2 minutes and a long cadence of 30 minutes that are collected into Mikulski Archive for Space Telescopes (MAST) \footnote{\url{https://archive.stsci.edu}} in both target pixels and light curve files \citep{Ricker2015, Sullivan2015, Feinstein2019}. Since a network map of a light curve needs many data points that are able to show the fine structure details, we develop the network approach for short cadence \tess\ observations. 
Short cadence light curves provide us with enough data points for making stable and reliable networks and they have Nyquist frequencies well above \dsct\ frequencies ranges (5-80\,d$^{-1}$). We collect our HADS star listed in Table \ref{tab1} from the literature and LADS stars are selected from the multimode \dsct\ stars studied in \citep{Hasanzadeh2021}, listed in Table \ref{tab2}. Then, we query the pre-search data conditioning (PDC) SAP flux of light curves for 31 HADS (Table \ref{tab1}) and 38 LADS (Table \ref{tab2}) stars, by using the Python Lightkurve package \citep{Lightkurve}. To correct systematic trends, the mean flux value is subtracted for each sector. Since the systematic correction tools (such as CBV-corrector) don't show significant changes in the results, we don't insist on using them. 

We use the gap-filled light curves for measuring the dependence of the average shortest path length with the size of the light curves (see section \ref{star network}). Since this metric is sensitive to the size of the light curves, we filled the gaps in order to factor out their effect for different lengths of light curves. Gaps appearing in the middle of each sector having a duration of about 1 day are caused by the downlink of \tess\ data to Earth for processing (see \tess\ Instrument Handbook from \citet{Vanderspek2018} for details). To avoid the gap effects, light curves are gap-filled by applying MIARMA\citep{PG15a, PG18}, which is an open-source\footnote{\href{https://github.com/javier-iaa/MIARMA}{https://github.com/javier-iaa/MIARMA}} code based on interpolation using ARMA models.
 Then, we performed tests on three different datasets from the same light curve: original data with gaps, gap-filled data, and cut light curve (a segment corresponding to the first 6527 data points with no gap). Our tests showed that gaps can affect degree distributions, but the cut light curves behave the same as gap-filled light curves (please see the comparative Figure \ref{figb1} in appendix \ref{B}). Therefore, we decided to use the cut light curves for our degree distribution analyses.
We also found that the average clustering shows a stable behaviour for the size of cut light curves, so we used the same-size cut light curves for the average clustering study (see Figure \ref{fig3}).

\section{Methods}\label{methods}
Complex network analysis is a helpful tool for studying the characteristics of natural complex systems such as \dsct\ stars that evolve via complex dynamics. In this section, we describe how the light curves of our HADS and LADS sample stars are mapped to horizontal visibility graphs and how the graph degree distributions are characterised by different distribution functions.

   \subsection{Network representation of a stellar light curve}\label{sec:HVG}
\begin{figure}
   \centering
    \includegraphics[width=0.49\textwidth]{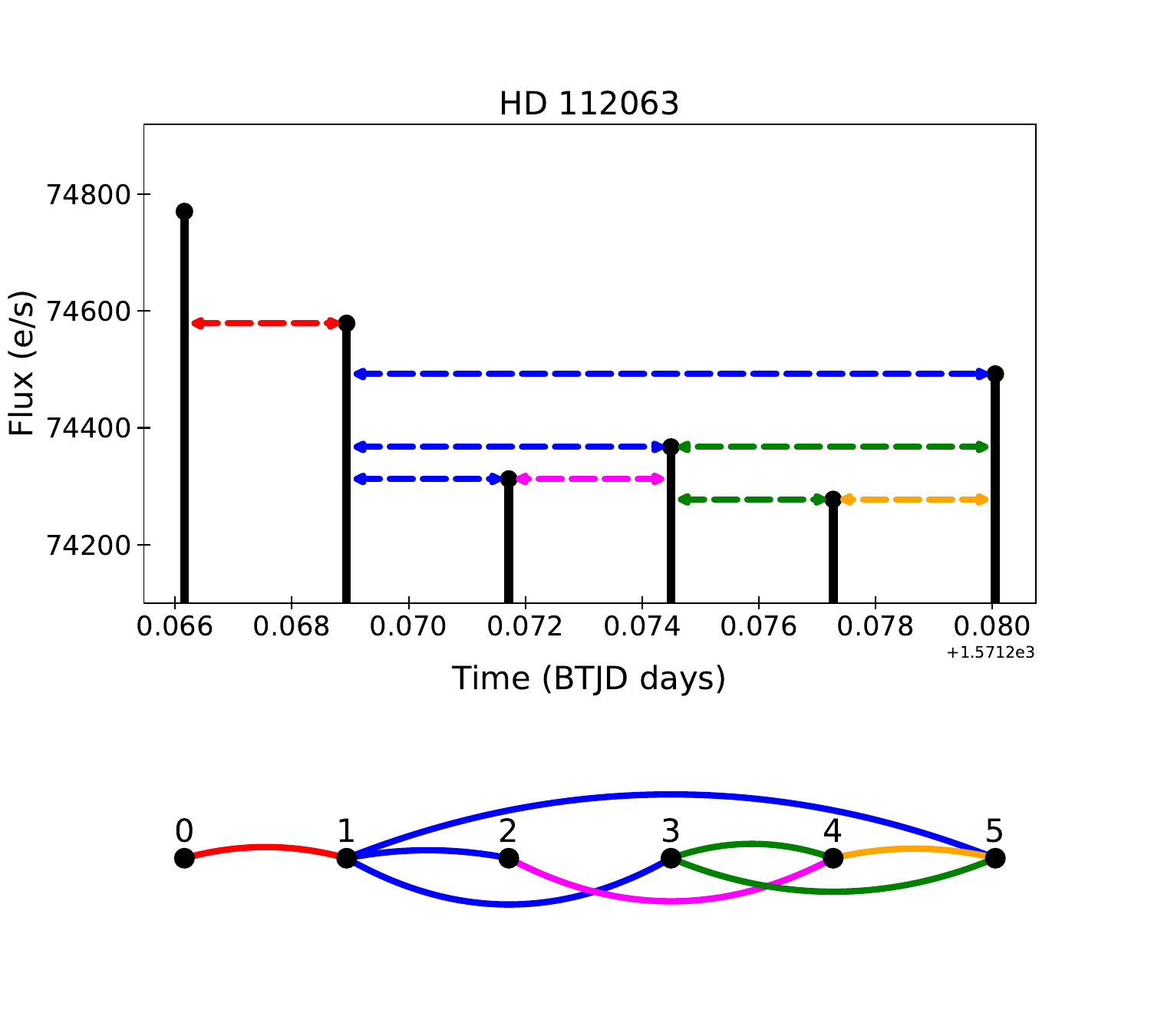}
   \caption{Horizontal visibilities for the first six points of the \tess\ light curve of HD 112063 (TIC 9591460) \dsct\ star when points can satisfy Equation (\ref{HVG}) (upper panel). The nodes are connected in the equivalent undirected HVG based on the horizontal visibility lines (bottom panel).}
   \label{fig1}  
   \end{figure}

   Horizontal visibility graph (HVG) that focuses on the interactions of system elements is one of the visibility algorithms for building a network for a time series \citep{lacasa2008, lacasa2010,luque2009horizontal}. To map a time series like a stellar light curve to a network (or a graph), via the horizontal visibility algorithm, each light curve point presents a node (vertex) in the equivalent graph, and only horizontal lines of sight between the times series points can determine the connection between the network nodes, then, nodes are connected in pairs by lines (edges or links). 
   Figure \ref{fig1} illustrates all the connections for the first six points of the light curve (first six nodes in the equivalent graph) through undirected HVG algorithms for HD 112063 (TIC 9591460).

   In an undirected HVG, node $(t_a, y_a)$ connects to $(t_b, y_b)$ if in the light curve an arbitrary point $(t_c, y_c)$ such that $t_a<t_c<t_b$ satisfies the condition
   \begin{equation}
   \centering
         y_a,  y_b > y_c.
         \label{HVG}
   \end{equation}

   In the network approach, we have networks (or graphs) to be considered instead of complex systems. A set of local and global metrics reflect particular features of these systems. Local metrics describe individual nodes or edges, and global metrics interpret the graph as a whole. We briefly define the node degree, clustering coefficient, shortest path length, and transitivity properties of an HVG network. 
   \begin{itemize}
   \item Degree of nodes: The node degree of a network is one of the local metrics that measures the centrality of each node. The degree of an individual node is defined as the number of edges linked to that node. Centrality shows the most influential nodes with effective connectivity through the network. In the current work we consider the distribution of all degree of nodes for every HVG of the stellar light curve.
   \item Average clustering: Local and average clustering coefficients are network metrics that indicate the tendency of nodes neighbours to be clustered. The local clustering coefficient of every single node measures the fraction of complete triangles to not-completed ones including the node. It is a number between 0 and 1, such that 0 means there is no connection between the node neighbours and 1 means all neighbours are fully connected. By taking an average of these local values, the average clustering is obtained.
   \item Transitivity: This is a global clustering coefficient that determines the density of triangles in a complex network. Transitivity is a global metric that measures the fraction of triples with their third edge serving to complete the triangle. 
   \item Average shortest path length: this path is another global metric for complex networks that defines the average number of links as the shortest paths for all pairs of nodes \citep{newman2003, boccaletti2006, newman2010, Acosta-Tripailao2021Entrp}.

   \end{itemize}
\subsection{Exponential distribution fits}\label{exp fits}
\citet{lacasa2010} showed that a given time series, independent of the generating probability function, maps to an HVG with an exponential degree distribution that can be applied to explore the dynamic nature of the complex system. In this method, the exponential index ($\lambda$) can indicate the stochastic or chaotic characterisation of the generating process of the time series. We fit an exponential function (Equation \ref{p_exp}) on the probability density function (PDF) of node degrees of the HVGs obtained from the light curves. 

\begin{equation}
   {\rm PDF}\simeq \exp(-\lambda k). \label{p_exp}
   \end{equation}

In HVG degree distributions described by the Equation \ref{p_exp}, indices $\lambda<\ln(3/2)$ are related to more chaotic systems while $\lambda>\ln(3/2)$ means that the system is governed by correlated stochastic processes which are less unpredictable. Indices equal to  $\ln(3/2)$  describes an uncorrelated random process time series \citep{lacasa2010,zou2019}. 

We examine a hypothesis test based on Kolmogorov–Smirnov test. The null hypothesis supposes no significant difference between the degree distribution of nodes and the exponential model. However, the alternative hypothesis assumes a substantial difference between the degree distribution of nodes and the model. We calculate a p-value to decide whether or not the exponential distribution hypothesis is compatible with the distributions. A p-value lower than the threshold of 0.05 rejects the null hypothesis (H0) showing that the exponential distribution is ruled out. We cannot deny H0 for a p-value higher than the threshold of 0.05. 

\subsection{Lognormal distribution fits}\label{log fits} 
   We also apply the maximum likelihood estimation to obtain the fit parameters of a lognormal model to the HVG degree distributions due to the relation between the lognormal distributions and multiplicative mechanisms. A lognormal distribution can describe the multiplicative process with independent varying parameters \citep{Bazarghan2008A&A, Farhang2022ApJ} that might happen in systems having fractal features \citep{Pietronero1986} such what has been studied in \dsct\ light curves \citep{deFranciscis2018MNRAS, deFranciscis2019MNRAS}.

   The lognormal probability density function for discrete variables is introduced by 

    \begin{equation}
   {\rm PDF}(k,\mu,\sigma)=\frac{1}{k A(\mu,\sigma)}\exp\left(-\frac{(\log k -\mu)^2}{2\sigma^2}\right)\label{equation2}
   \end{equation}
   with, 
   \begin{equation}
   A(\mu,\sigma)=\sum_{k=1}^{\infty}\frac{1}{k}\exp\left(-\frac{(\log k -\mu)^2}{2\sigma^2}\right).
   \end{equation}

where $\mu$ and $\sigma$ are the lognormal parameters. 
By examining a hypothesis test based on Kolmogorov–Smirnov test, we cannot reject H0 for a p-value higher than the threshold of 0.05. 

\section{Results and discussions}\label{results}
The complexity characteristics of \dsct\ stars can be investigated by network analysis. We apply the HVG algorithms to investigate the characteristics of 31 HADS (Table \ref{tab1}) and 38 LADS (Table \ref{tab2}), respectively.

\subsection{\dsct\ networks}\label{star network}

\begin{figure}[b]
\centering
\includegraphics[width=6.8cm,height=7.6cm]{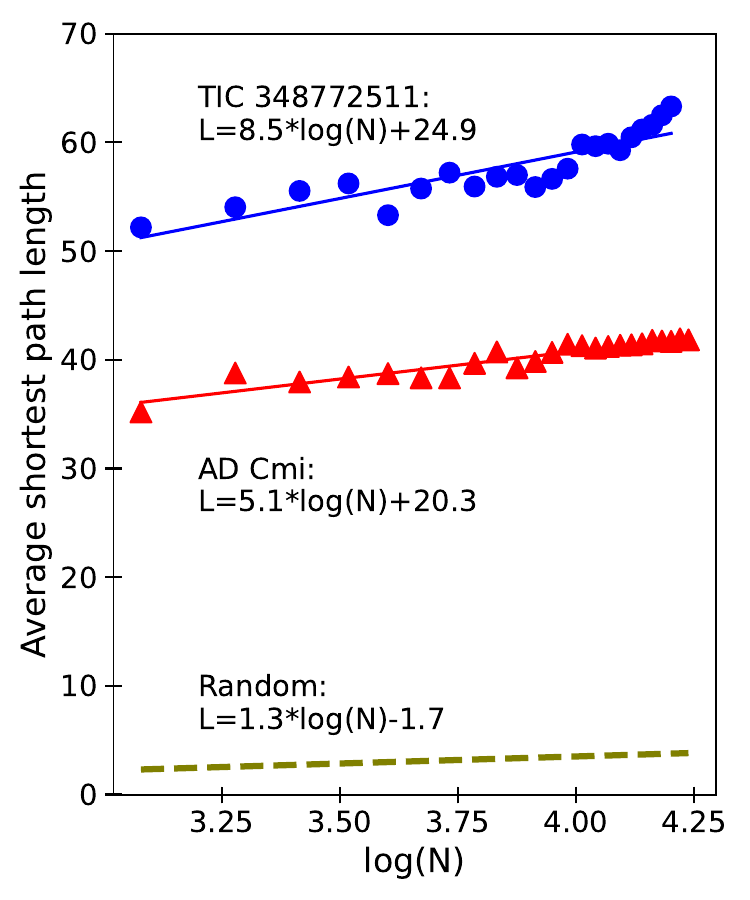}
\caption{The average shortest path length of the HVGs versus the logarithm of light curve sizes for TIC 348772511 or HD 21190 as a LADS star, AD Cmi as a HADS star, and an equivalent random network indicated by blue circles, red triangles, and olive dashed line respectively. The fitted line for HD 21190 (blue line) and AD Cmi star (red line) deviates from the random network.}
\label{fig2}
\end{figure}

Figure \ref{fig2} represents the average shortest path length for the HVG of the gap-filled light curves for AD Cmi (HADS) and HD 21190 or TIC 348772511 (LADS) compared with the random network. Figure \ref{fig2} shows that the average shortest path length of both stars has significantly deviated from the average shortest path length of a random network. \citet{luque2009horizontal} proved that the HVG's average shortest path length for any random time series is independent of the generating probability function and has a logarithmic dependence with the number of time series data points (N): $1.3\ log(N)-1.7$. The linear dependency of the average shortest path length to the logarithm of light curve size for both cases indicated the small-world behaviour of networks \citep{Watts-nature1998}. We observe similar behaviour for all target stars of Tables \ref{tab1} and \ref{tab2}. The small-world behaviour for \dsct\ stars network shows that the high peaks at the light curve are connected to several neighbouring small peaks and the other high peaks at the light curve. 
In the context of complex networks, small-world networks have been explored in several studies \citep{Watts-nature1998, Mathias-PRE, Latora-PRL}. 

\begin{figure}
\centering

\tabskip=0pt
\valign{#\cr
  \hbox{%
    \begin{subfigure}
    \centering
    \includegraphics[width=6.8cm,height=6.10cm]{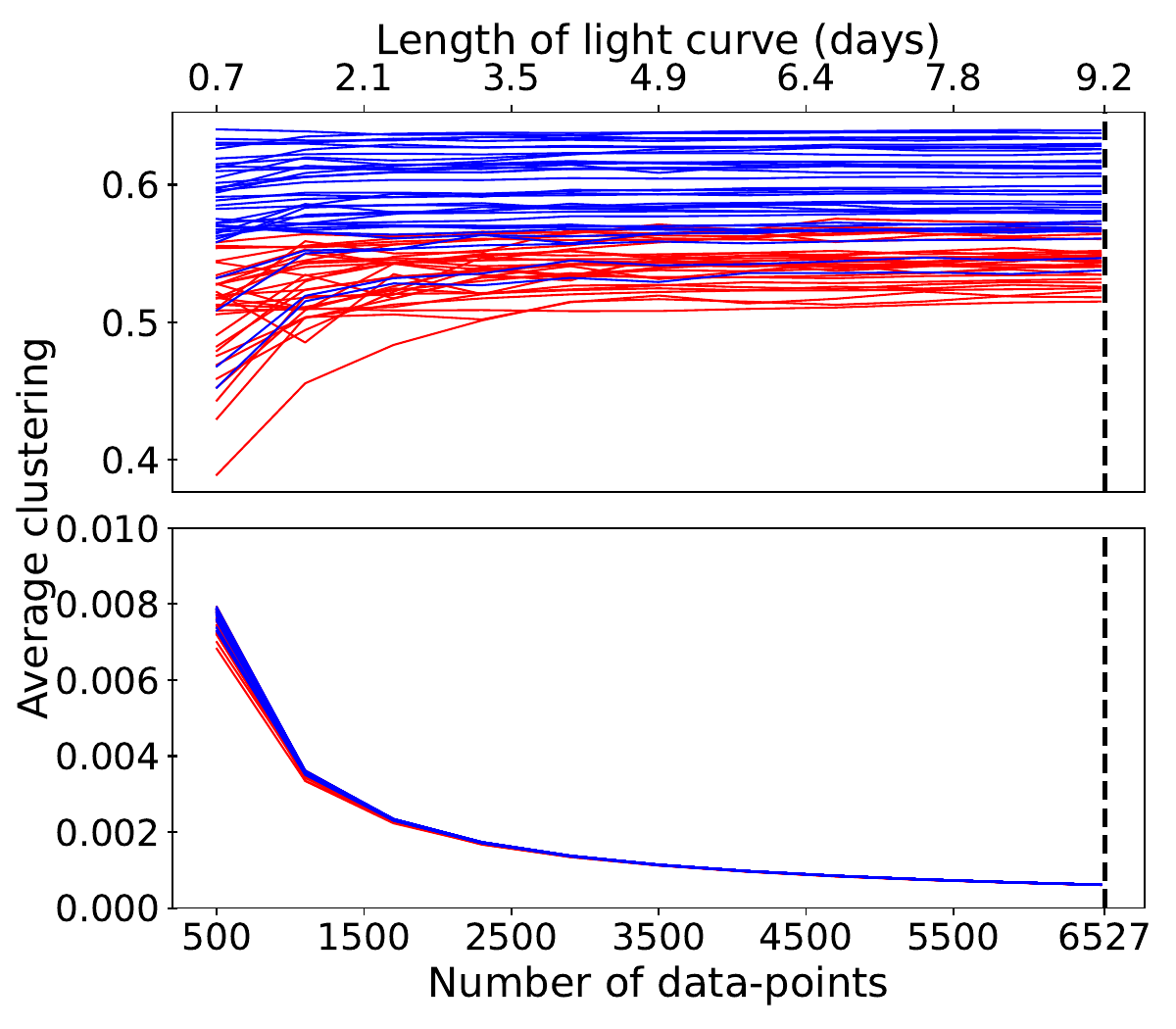}%

    \end{subfigure}%
  }\cr
  \noalign{\hfill}
  \hbox{%
    \begin{subfigure}
    \centering
    \includegraphics[width=2.20cm,height=2.15cm]{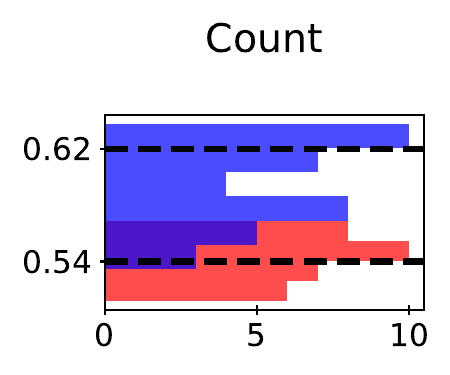}%

    \end{subfigure}%
  }\cr
}
\caption{The HVG average clustering coefficients versus the graph sizes for HADS (red lines), LADS (blue lines) stars in up-left panel, and for their equivalent random networks in down-left panel. The distribution of average clustering coefficients at the size of 6527 nodes shows different maximum values for HADS (0.62) and LADS (0.54) in the up-right panel.}
\label{fig3}
\end{figure}

Figure \ref{fig3} depicts the average clustering coefficients of HVGs versus the light curve sizes for HADS (red lines), LADS (blue lines) stars, and also for their equivalent random networks which are small-worlds. The average clustering coefficient changes for small light curve sizes until a quasi-stable network is formed, and then the average clustering coefficient remains approximately constant. We cut our light curve at 6527 data points (or 9.2 days) which is shown by a black dashed line where the average clustering property of networks seems stable and this light curve size can be used for our analysis. The size of 6527 data points (or 9.2 days) comes from the smallest segment size among our 69 light curves that doesn't have any gap. By considering the distribution of average clustering coefficients at the cut-line, the HVG average clustering coefficients of LADS accumulate around 0.62 which is slightly greater than 0.54 as the peak of average clustering coefficients for HADS, and all stars are considerably separated from equivalent random networks. The average clustering for equivalent small-world random networks ($C_R$) is calculated by measuring the average node degree of each network, then divided by the number of network nodes ($C_R=$ average degree of nodes / number of nodes) \citep{Watts-nature1998, Gheibi2017ApJ}

\begin{figure}
         \centering
         \includegraphics[width=0.470\textwidth]
         {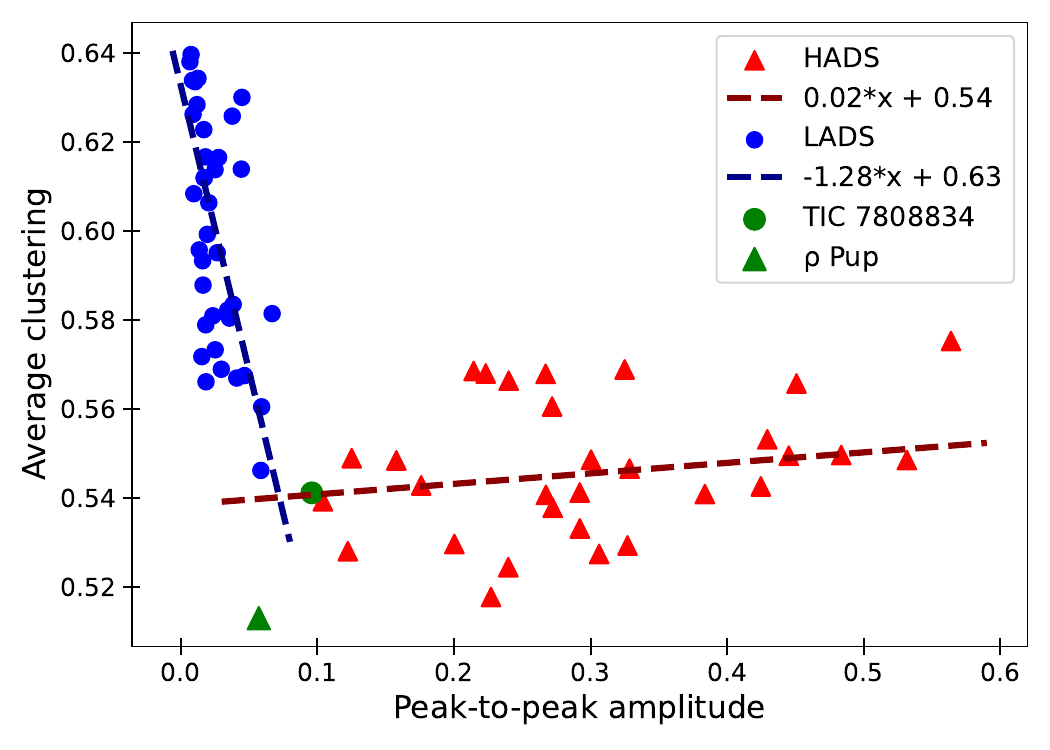} 
         \caption{Scatter plot of HVG average clustering versus the peak-to-peak amplitude of \tess\ light curves can separate the HADS (red triangles) and LADS (blue circles) in two clusters. Two linear functions (dashed lines) describe the dependency of Average clustering and peak-to-peak amplitude values as morphological parameters in network and light curve respectively.}
\label{fig4}
\end{figure}

Figure \ref{fig4} presents the dependency of HVGs average clustering coefficients on the peak-to-peak amplitude of \tess\ light curves. The variety of average clustering coefficients between the same-size networks of LADS and HADS is differentiable by considering the peak-to-peak amplitudes in Figure \ref{fig4}. LADS stars (blue circles) occupy higher average clustering values with a tiny range of low peak-to-peak amplitudes but HADS stars (red triangles) occupy lower average clustering values with higher peak-to-peak amplitudes. Average clustering is a parameter defined in the time domain which is closely related to the presence of a fine structure in the light curves, i.e. roughness. This fine structure can only be represented by a high number of frequency components \citep[for example][]{IACOBELLO20181}. In Figure \ref{fig4}, the dependence of average clustering with the peak-to-peak amplitudes shows that LADS light curves generally have a higher average clustering coefficient and it decreases for higher amplitudes. Since pulsation mode energies are also higher for higher peak-to-peak amplitude, this evidence points to a transition from high to low energies injected by the star in the oscillations favouring a few high amplitude radial modes (low average clustering in HADS) in contrast to a high number of low amplitude non-radial modes (high average clustering in LADS). 
The green circle is a LADS star (TIC 7808834) that is located nearer the transition between high clustering (LADS) and low clustering (HADS). For that, this target in particular deserves an in-depth frequency analysis. Furthermore, the green triangle shows $\rho$ Pup which has a mono-mode HADS properties in the Fourier parameters but its amplitude in different filters is smaller than what is expected for HADS \citep{Nardetto2014}. The current analysis with average clustering coefficient can confirm $\rho$ Pup has HADS properties.

\begin{figure}
\centering
    \begin{subfigure}
    \centering
    \includegraphics[width=1.\linewidth]{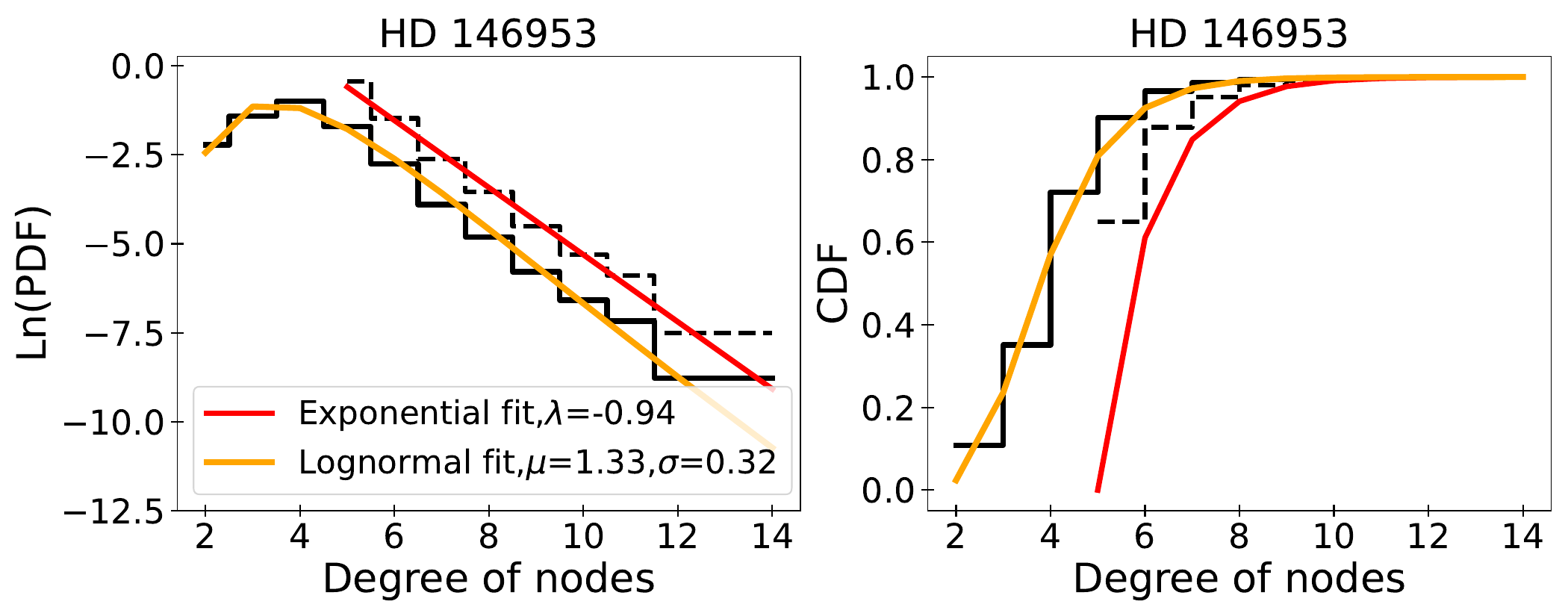}
     \end{subfigure}\\

     \begin{subfigure}
     \centering
     \includegraphics[width=1.\linewidth]{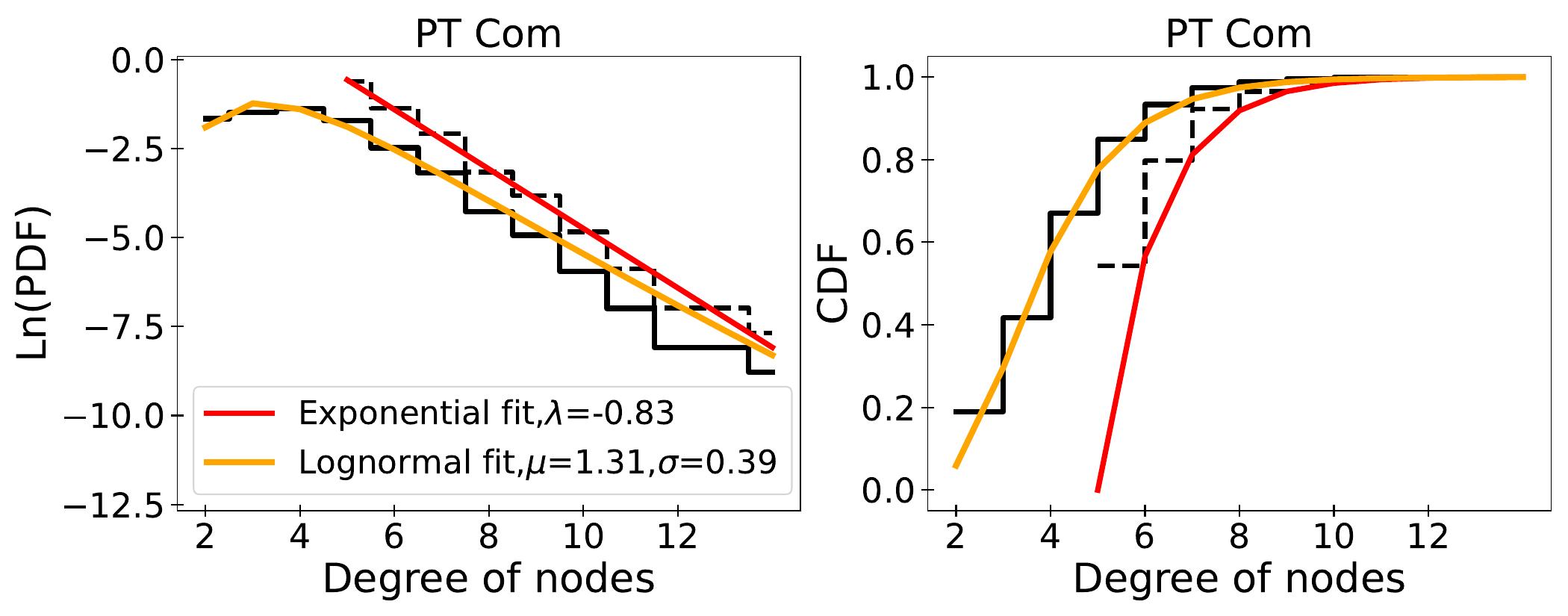}
     \end{subfigure}\\

     \begin{subfigure}
     \centering
     \includegraphics[width=1.\linewidth]{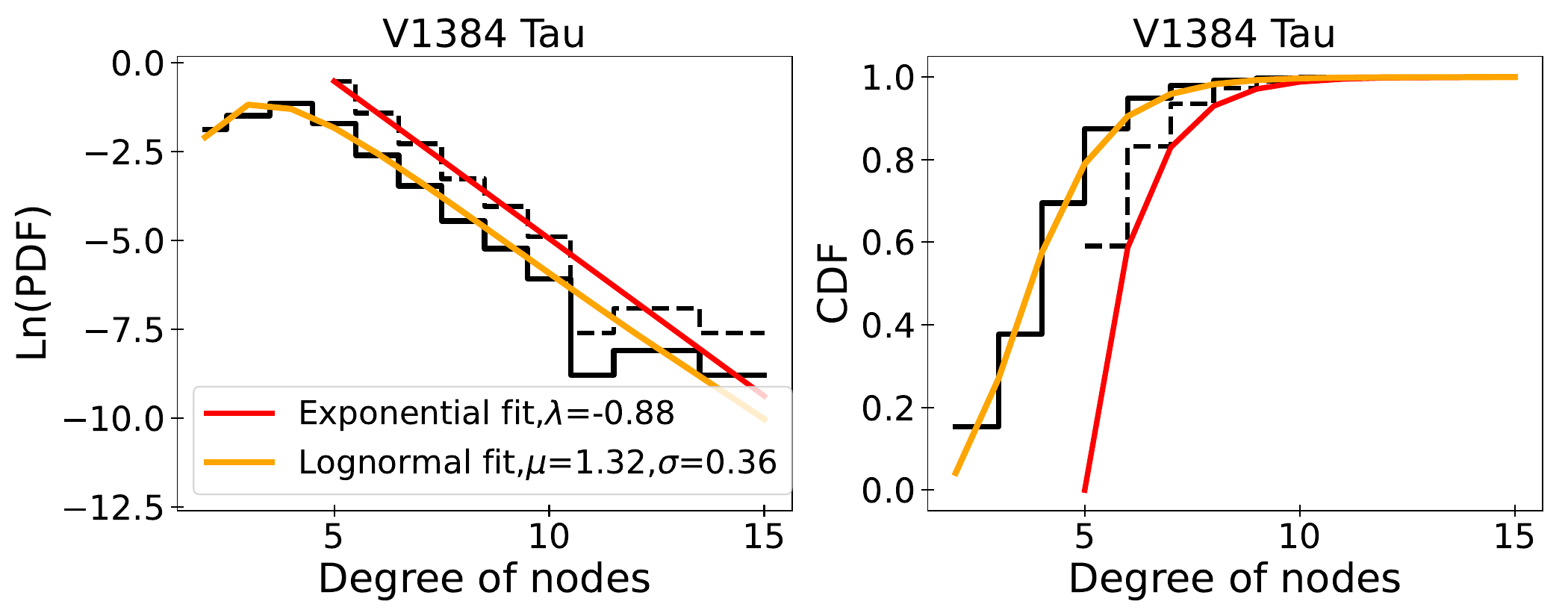}
     \end{subfigure}\\

      \caption{Logarithm of PDF (first column) and CDF (second column) for the HVGs nodes degree for three HADS (HD 146953, PT Com, and V1384 Tau) stars. Exponential fits (red lines) and lognormal distributions (orange lines) are plotted on their normalized distributions in solid and dashed lines, respectively.}
\label{fig5}
\end{figure}

\begin{figure}
     \centering
     \begin{subfigure}
         \centering
         \includegraphics[width=1.0\linewidth]{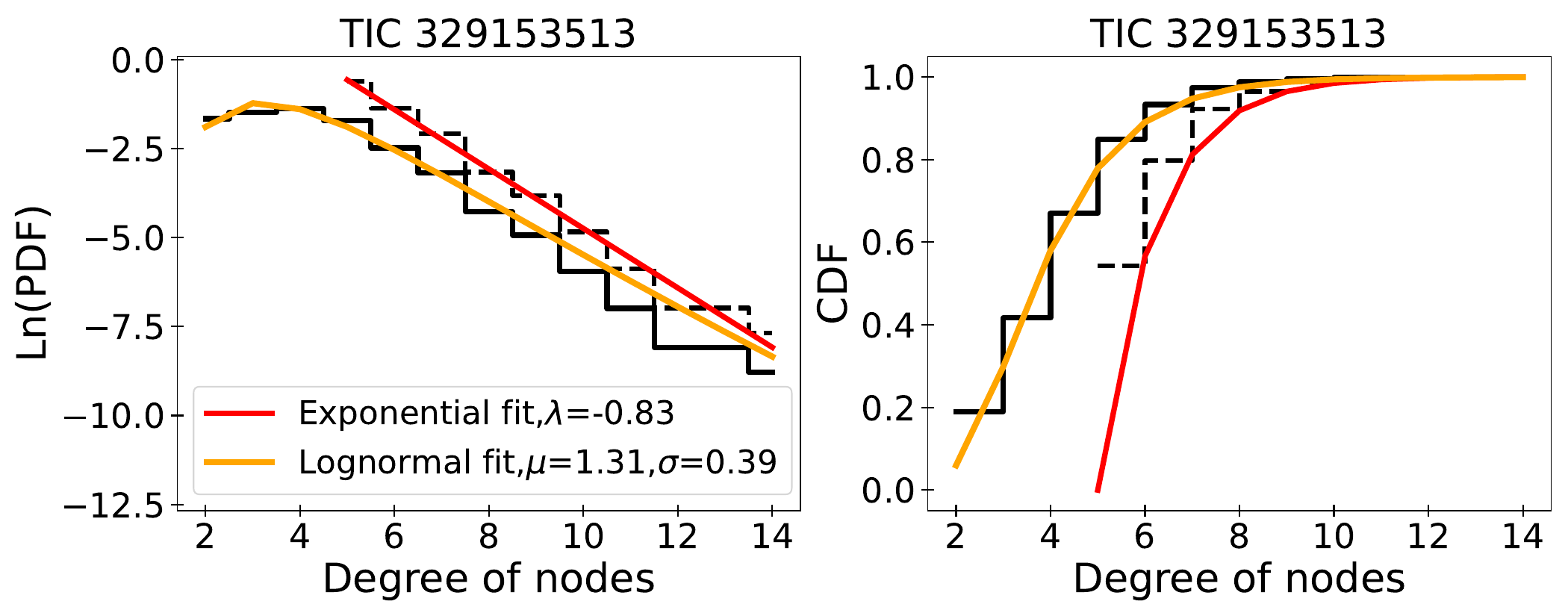}
     \end{subfigure}\\

     \begin{subfigure}
         \centering
         \includegraphics[width=1.0\linewidth]{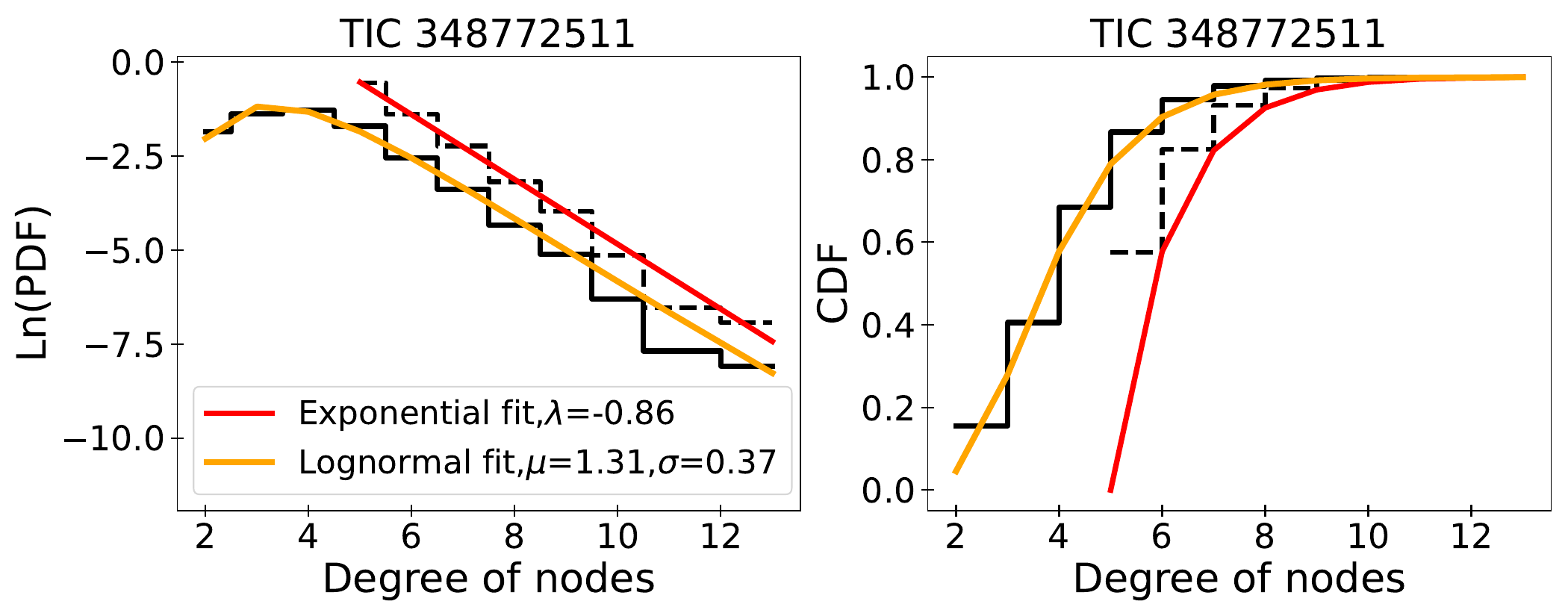}
      \end{subfigure}\\

     \begin{subfigure}
         \centering
         \includegraphics[width=1.0\linewidth]{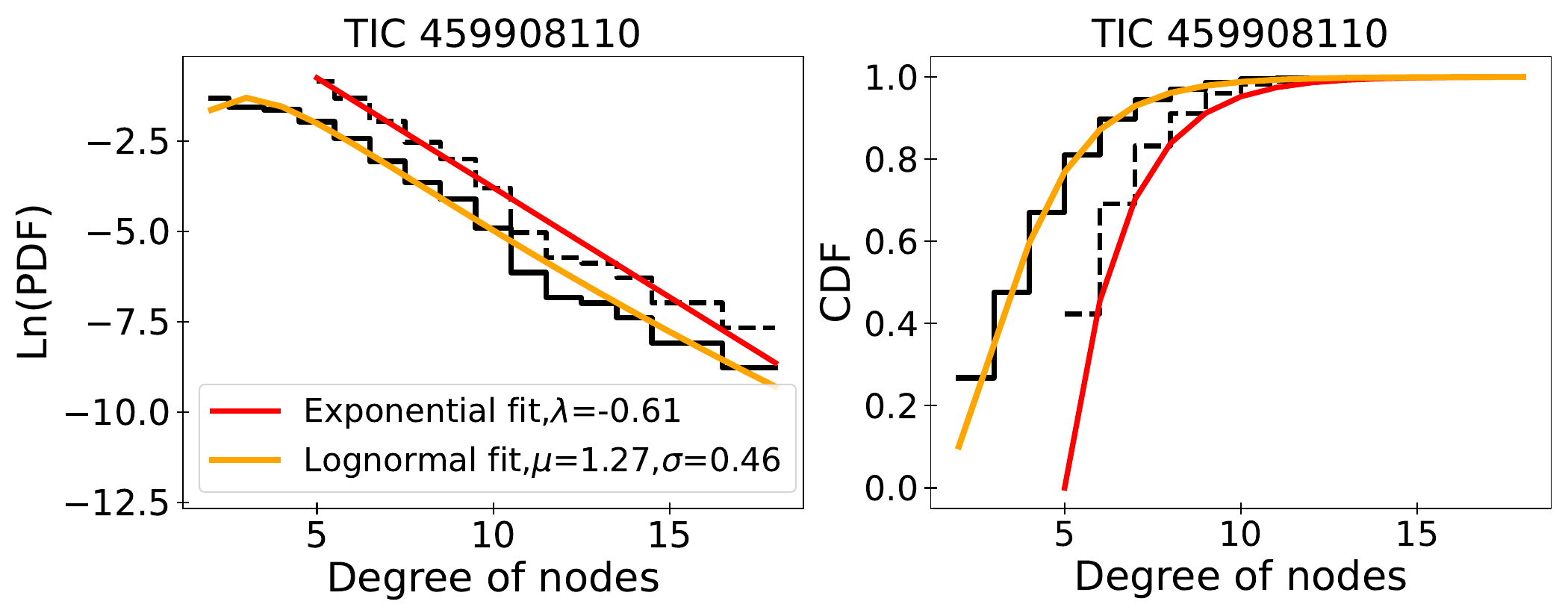}
     \end{subfigure}\\

      \caption{Logarithm of PDF (first column) and CDF (second column) for the HVGs nodes degree for three LADS (TIC 329153513, TIC 348772511, and TIC 459908110) stars. Exponential fits (red lines) and lognormal distributions (orange lines) are plotted on their normalized distributions in solid and dashed lines, respectively.}
\label{fig6}
\end{figure}

Figures \ref{fig5} and \ref{fig6} show PDFs and CDFs of degrees for HVG nodes of three HADS and three LADS stars, respectively. We fit the exponential distribution described in section \ref{exp fits}, to the HVG degree distributions in red lines with p-values quite higher than 0.05. 
Our analysis shows the robustness of exponential fits in the framework of the study \cite{lacasa2010}, giving us several hints on the physical meaning for distributions and the value of indices. 
Exponential indices can distinguish between the stochastic or chaotic nature of the dynamics that generate the stellar light curve as people investigate in other fields of study \citep{Ganesh2020,BRAGA2016}. Concerning the corresponding spatial processes of energy accumulation and discharge, exponential distributions have been found in time series for grain density of different Self-organising Criticality (SOC) sand-pile models \citep{Kaki22}. By applying a robust linear regression on the logarithm of the PDF for HVG node degrees that allows two fit parameters as the slope and the intercept, we obtain slopes higher than $\lambda=\ln(3/2)$. These exponential indices higher than $\ln(3/2)$=0.41 might mean that the stellar brightness variations are governed by the correlated stochastic process. We calculate exponential fits for the range of degrees $k \in [4,+\infty]$, with the corresponding normalization (black dashed line PDF of node degrees in Figure \ref{fig5} and \ref{fig6}). Please see Tables \ref{tab1} and \ref{tab2} for exponential indices, uncertanities and p-values.

In addition, we fit the PDFs and CDFs of degrees for HVG nodes of light curves with lognormal models explained in section \ref{log fits}, obtaining high p-values and two distribution parameters with negligible uncertainties (Tables \ref{tab1} and \ref{tab2}). In Figures \ref{fig5} and \ref{fig6} the lognormal distributions are plotted in orange lines.
Let us remind that a lognormal distribution is induced by some multiplicative stochastic process, e.g., a product of many independent stochastic variables, $Q_{T} = \prod_{t=1}^{T}q_{t}$. Its logarithm is a sum of many independent stochastic variables, $ln(Q_{T}) =\sum_{t=1}^{T} ln(q_{t})$, which is Gaussian distributed. Using a heuristic argument we observe that if node $t$ in the HVG is linked to node $t+\tau$, then in the corresponding time series, the height $h$ of all nodes $\left\{t+1,\dots,t+\tau-1\right\}$ has to be lower than the height of $t$ and $t+\tau$. So the probability $P(t\rightarrow t+\tau)$ to have a link in the network between $t$ and $t+\tau$ depends conditionally on the product $\prod_{\bar{t}=t}^{t+\tau}P_{\bar{t}}$ of suitable probability functions $P_{\bar{t}}(h_{\bar{t}}<h_{t},h_{t+\tau})$ of all the heights in between, i.e. on the whole time story of the signal, whose values may follow distributions from independent stochastic variables. This probability $P(t\rightarrow t+\tau)$ reminds a multiplicative process. Observationally, lognormal distributions have been obtained in some studies of different energetic surface phenomena in solar physics \citep[for example][]{Alipour2022A&A, Farhang2022ApJ}.
We could not discard neither a power law behaviour for high values of degree, since in some conditions a lognormal distribution could switch to a power law distribution  \citep{mitzenmacher2004}. Following the product of probabilities argument of the former paragraph, for some specific distributions of height, PDF result to be the sum of lognormal distributions, which would have essentially a lognormal body but a power law distribution in the tail. Even in presence of a single lognormal if $\sigma$ is an appreciable amount of the range of degree we are observing, one can confound lognormal with power law. Finally a small change in the constraints of multiplicative process allows a power law distribution. As long as there is a bounded minimum that acts as a lower reflective barrier to the multiplicative model, it will yield a power law instead of a lognormal distribution. Power law distributions have been measured in models and observation of energy discharges process, like in VG for number of sunspots vs time \cite{Zou14}, as well as HVG for avalanches size in the classical Self-organised Criticality BTW model \cite{Adami24}. 

\subsection{White noise signature in the HVG networks}
\citet{luque2009horizontal} showed that the degree of nodes ($k\geq 2$) for the HVG map of an uncorrelated random time series follows the theoretical exponential distribution of Equation \ref{pdf_r} which can be obtained from Equation \ref{p_exp} when $\lambda=\ln(3/2)$.

\begin{equation}
    \rm PDF = \frac{1}{3}\left(\frac{2}{3}\right)^{k-2}.\label{pdf_r}
\end{equation} 

  They also showed that this theoretical expression is independent of the probability distribution from which the random time series was created. Using Equation (\ref{pdf_r}), we obtain the theoretical cumulative distribution function (CDF$_T$) as
\begin{equation}
{\rm CDF}_T(k) = \frac{1}{3}\left(\ln\frac{2}{3}\right)^{-1}\left(\frac{2}{3}\right)^{-1}\left(\left(\frac{2}{3}\right)^{k-1}- 1\right).\label{cdf}
\end{equation} 

The complementary cumulative distribution function (CCDF) is useful for describing the tail distributions and is related to CDF as CCDF=1-CDF. 

\begin{figure}
     \centering
     \includegraphics[width=6.4cm,height=9.1cm]
     {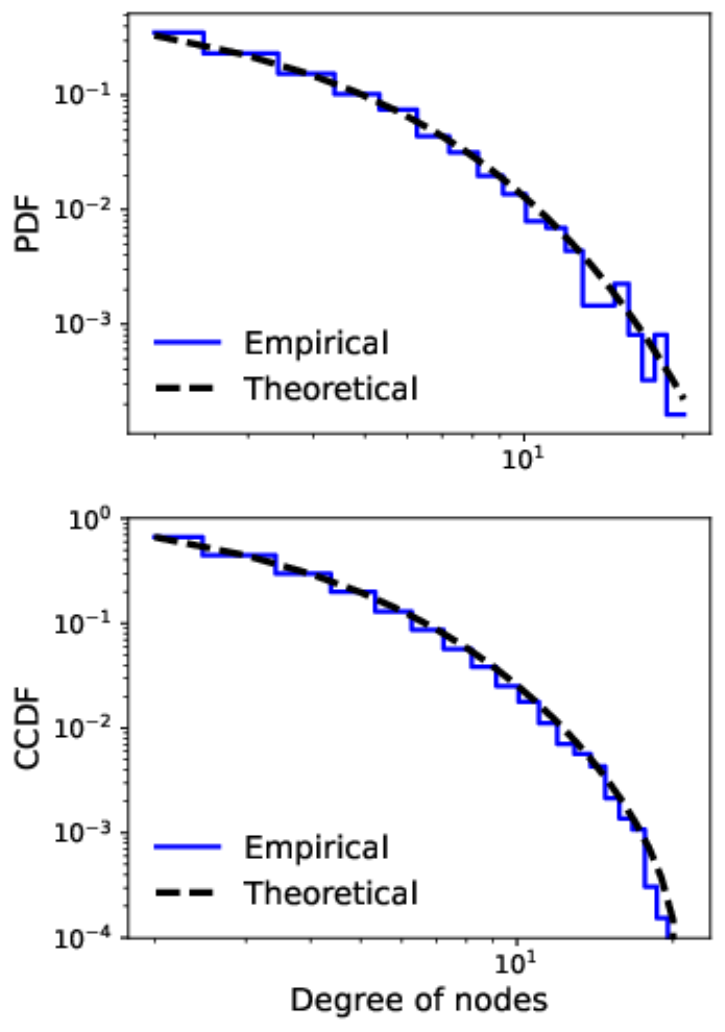}  
     \centering
     \caption{PDF (upper panel) and CCDF (bottom panel) of the degrees of the HVG nodes of a Gaussian random time series with 6527 data points (blue line) and the theoretical functions expressed in equations \ref{pdf_r} and \ref{cdf} respectively (black dashed line).}
\label{fig7}
\end{figure}

Figure \ref{fig7} represents the PDF (top panel) and the CCDF (bottom panel) of the degrees of the HVG nodes of a Gaussian random noise with 6527 data points (the same size as our light curves) and the theoretical expression (Equations \ref{pdf_r} and \ref{cdf}). We expect the HVG degree distribution for any random time series with arbitrarily generated probability distribution would excellently match the theoretical expression. The Kolmogorov-Smirnov distance ($KS_D$) is a measure to quantify the distance between distributions. The $KS_D$ is given by $KS_D={\rm sup}(|CDF_{\rm data}-CDF_{T}|)$, where $CDF_{\rm data}$ is the empirical cumulative distribution function \citep{mahmoud2000sorting} for the HVG degree distribution of time series. The $KS_D$ is distributed around 0.003 for various random time series with a size of 6527 data points. We suggest $KS_D$=0.003 as a criterion to distinguish pure Gaussian white noise with a size of 6527 data points. 

\begin{figure}
         \centering
         \includegraphics[width=8.4cm,height=5.8cm]{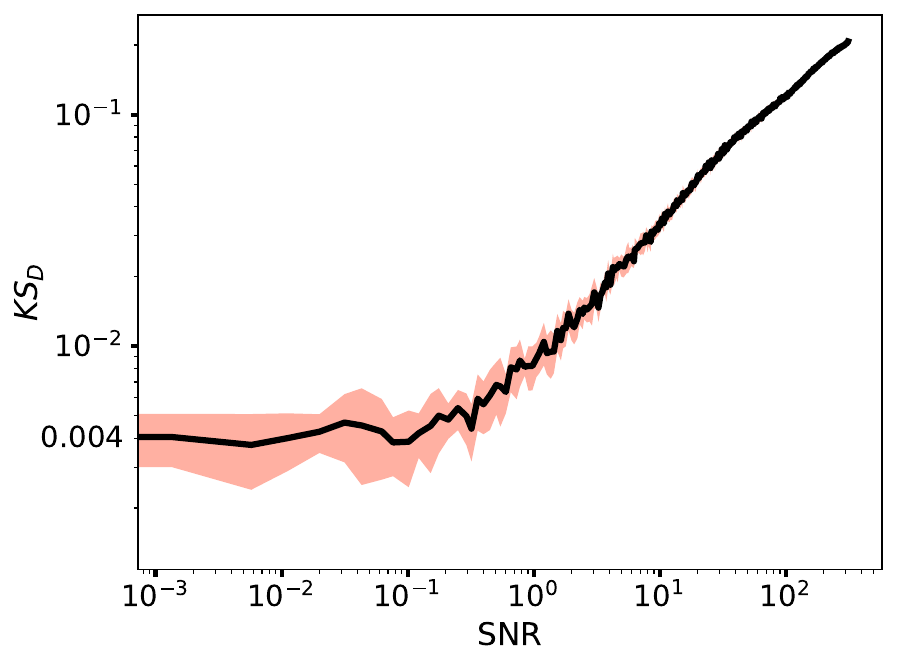}  
      \caption{The $KS_D$ values versus the signal-to-noise ratio (SNR) for the HVG degree distributions of synthetic time series with a size of 6527 data points. The red-shaded area shows the standard deviation of $KS_D$ for several white noises.}
\label{fig8}
\end{figure}

To figure out the role of noise on the HVG parameters of a light curve, we simulate a synthetic single-frequency time series with a size of 6527 data points in the presence of white noise with a variety of signal-to-noise ratio (SNR). The zero SNR corresponds to a zero-amplitude signal with random Gaussian white noise (such as Figure \ref{fig2}). Using the power of signal ($P_s$) and noise ($P_n$), the SNR is given by
   \begin{equation}
   {\rm SNR}=\frac{p_s-p_n}{p_n}.
   \end{equation}
Figure \ref{fig8} shows the $KS_D$ values for the HVG degree distributions of this synthetic noisy time series with various SNRs. The $KS_D$ is about 0.0040$\pm$0.0014 for tiny SNRs and increases with increasing SNR. Increasing the SNR slightly more than 0.1, the $KS_D$ is significantly higher than the noise network. The lower SNRs corresponded to the lower oscillation amplitudes in the noisy background. Therefore, $KS_D$ of the HVG degree distribution of time series or light curves may be a valuable criterion for the frequency pre-whitening process. In other words, if the HVG degree distribution for a light curve residual is matched with the theoretical expression for an uncorrelated random time series, the frequency cleaning is expected to stop (see also section \ref{star network}).  

\begin{figure}[h]
\centering
 \includegraphics[width=.50\linewidth]{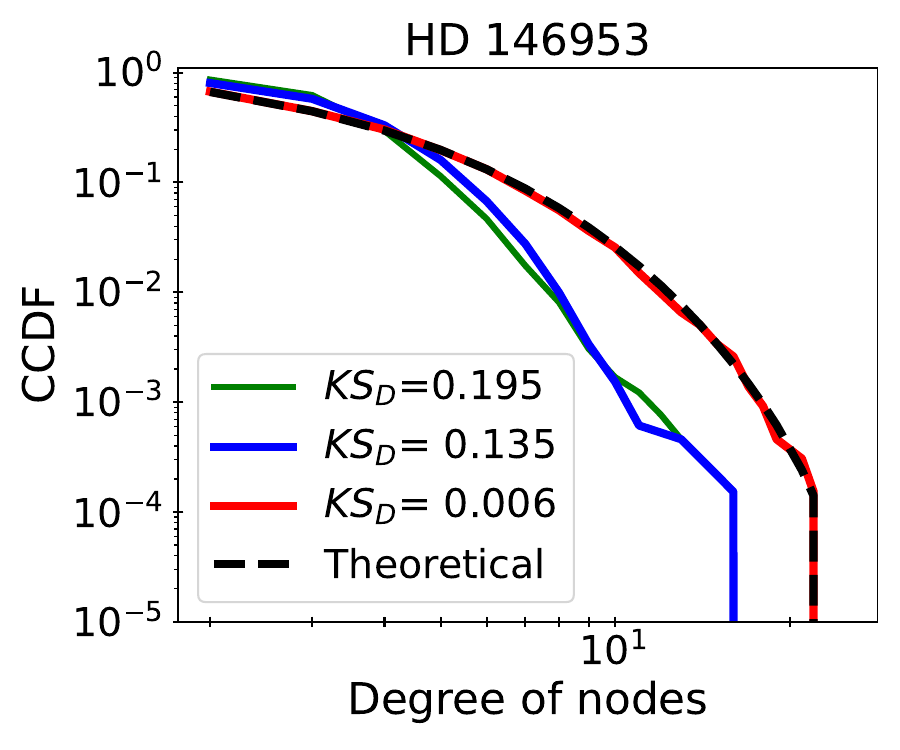}\hfill
  \includegraphics[width=.50\linewidth]{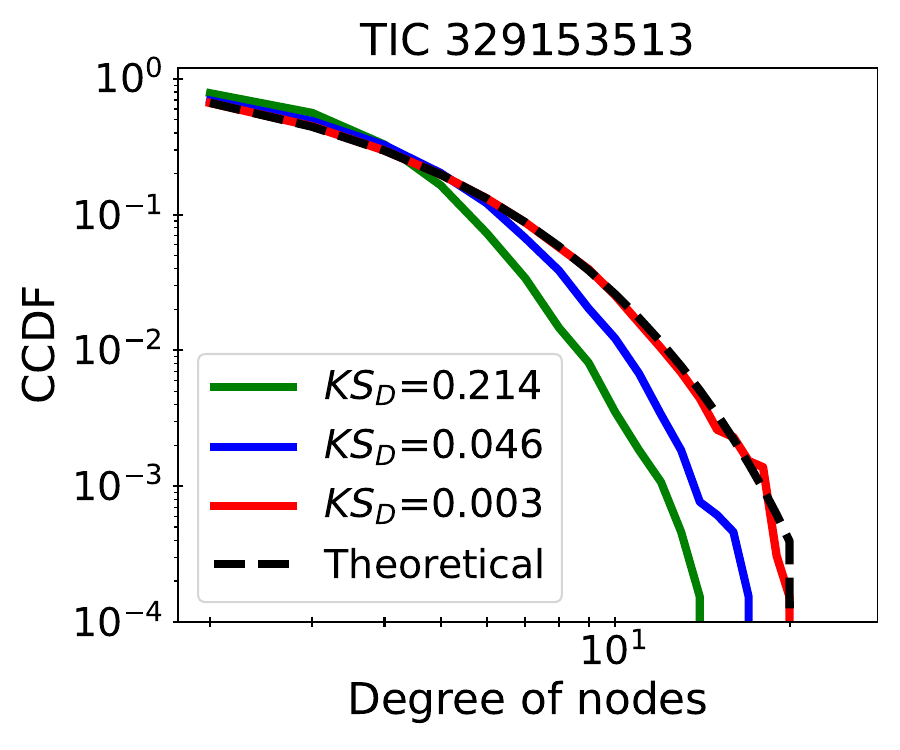}\hfill
  \includegraphics[width=.50\linewidth]{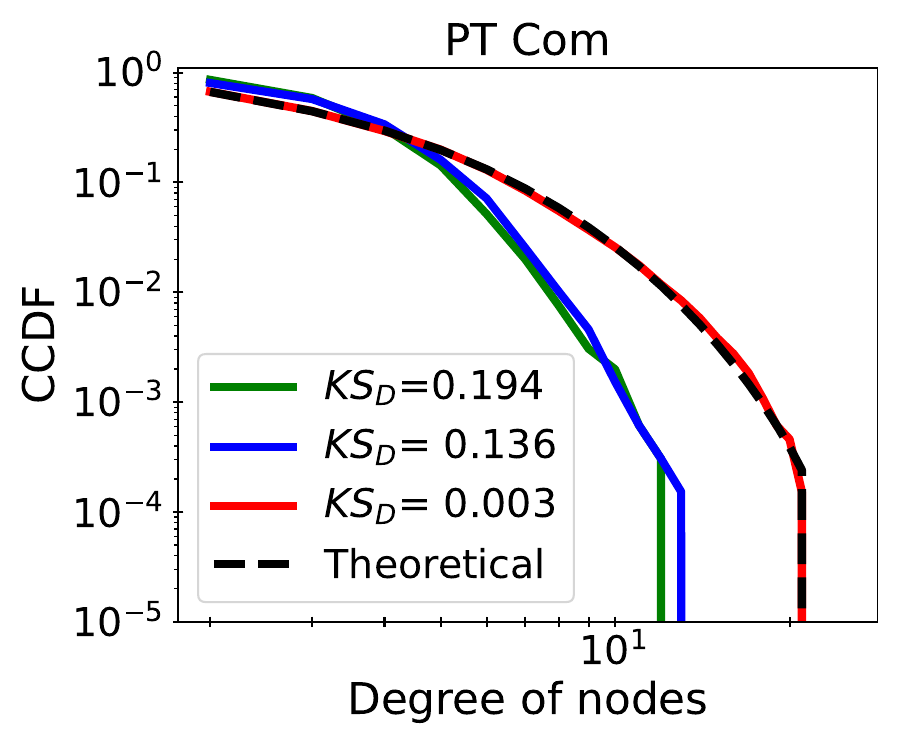}\hfill
  \includegraphics[width=.50\linewidth]{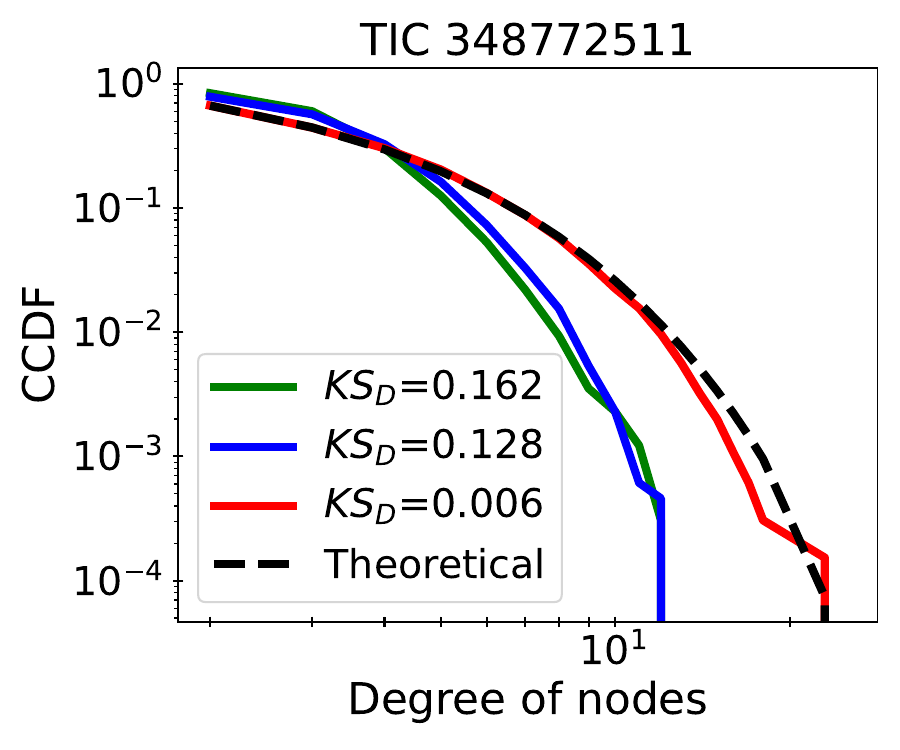}\hfill
    \includegraphics[width=.50\linewidth]{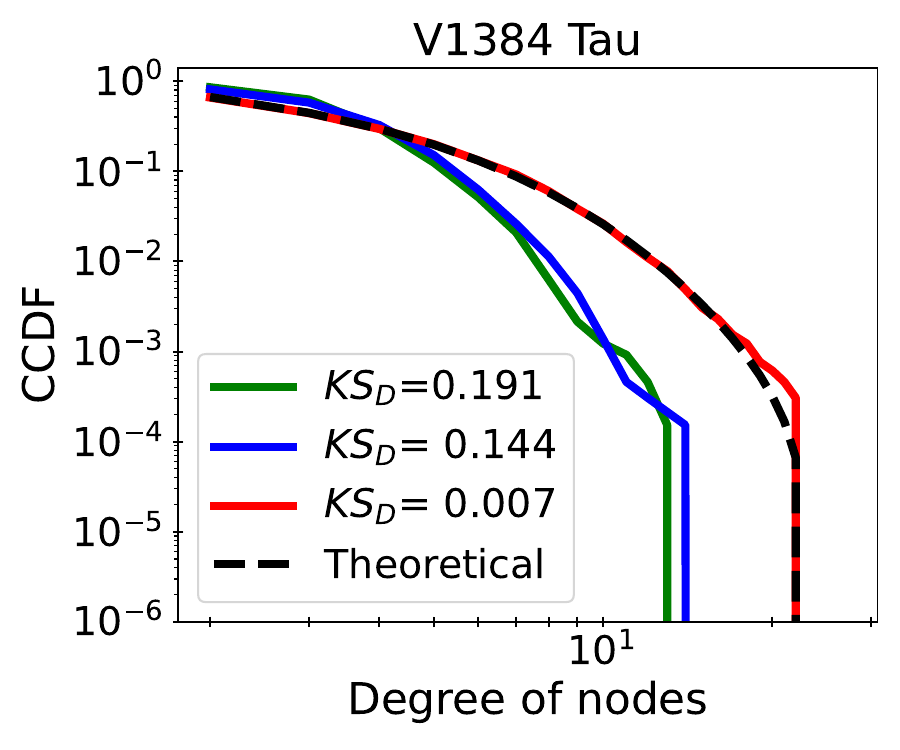}\hfill
    \includegraphics[width=.50\linewidth]{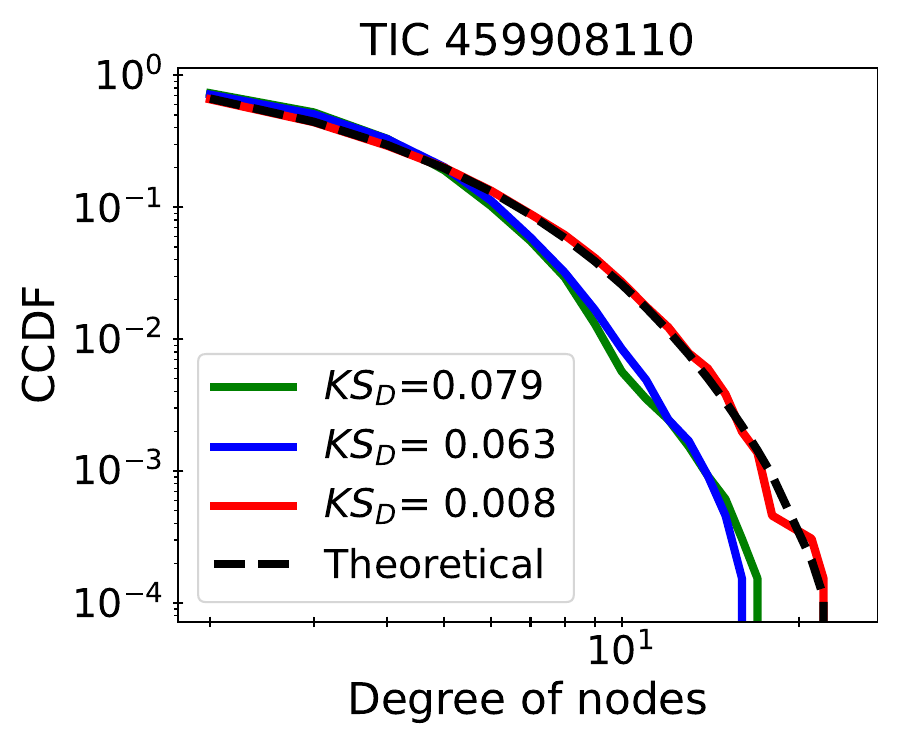}\hfill

   \caption{The HVG degree distributions of three HADS light curves (HD 146953, PT Com, V1384 Tau) in the left column and three LADS light curves (TIC 329153513, TIC 348772511, and TIC 459908110) in the right column, in presence of all frequencies (green), in absence of the dominant frequency (blue), and in the absence of all significant frequencies (red). The black dashed line shows the theoretical expression for random series (see Eq.\ref{cdf}). The $KS_D$ is presented for these three stages of pre-whitening for each star. Note that it is getting smaller, and quite close to the value of 0.0040$\pm$0.0014.}\label{fig9}
\end{figure}

Figure \ref{fig9} displays the CCDF of the degree nodes of the HVG networks for the same HADS star of Figure \ref{fig5} and LADS stars of Figure \ref{fig6} in the left and right columns respectively. The Kolmogorov–Smirnov distance $KS_D$ between the HVG degree distribution of random model (black dashed line) of Equation (\ref{cdf}) and the HVG degree distribution of pre-whitened residuals for light curve sizes of 6527 data points were computed. The pre-whitened residuals were obtained by using SigSpec \citep{Peter2012} where only the dominant frequency is removed (blue line) and all significant frequencies with SNR>4 are removed (red line). The green distribution corresponds to the degree distribution of the HVG of the original light curves where all frequencies are present. After removing all mode frequencies (SNR>4), for the noisy residual of HD 146953, PT Com, and V1384 Tau, the $KS_D$ values of 0.006, 0.003, and 0.007 were obtained, respectively. Also, the $KS_D$ values for TIC 329153513, TIC 384772511, and TIC 459908110 were calculated as 0.003, 0.006 and 0.008, respectively. These results imply that the highest peak frequency is not the only mode for these HADS stars, and reducing most of the oscillation modes from the \dsct\ light curves gives a mostly noisy residual that approximately obeys the degree of nodes of an uncorrelated random time series. 

The $KS_D$ about 0.004$\pm$0.0014 would be a helpful criterion for verifying noisy (white) residuals by removing oscillations of a pulsating star light curve. In the next step, this method will be developed for more stars, and a network approach is applied to recognize the coloured noise (e.g., red) in the stellar residuals.

\section{Conclusions}\label{concs}
We investigate the network characteristics for \dsct\ stars as non-linear complex systems. We focus on the undirected HVG's local and global metrics to discuss the properties of HADS and LADS stars (69 sample stars in Table \ref{tab1} and \ref{tab2}). 
\begin{itemize}
 
    \item The linear dependency of the HVG average shortest path length to the logarithm of the size of \dsct\ light curves (Figure \ref{fig2}). This behaviour for \dsct\ stars indicates the small-world complex networks with non-random properties. This small-world property implies that the peaks of the stellar flux light curves connect with some small close peaks and are then linked to other significant peaks along the light curves.       
    
    \item The average clustering coefficients of the HVGs for \dsct\ stars deviated from random models and also they are different for HADS and LADS (Figure \ref{fig3}). Scattering of average clustering coefficient and peak-to-peak amplitude of HADS and LADS stars can separate two \dsct\ subclasses in overlapping two clusters (Figure \ref{fig4}) that follow two linear behaviours, in general. The lower clustering coefficient for most HADS stars indicates the simpler light curve (containing one, two, or three independent modes then having a smoother light curve). In contrast, the higher clustering for LADS cases is a signature of the more complicated and rough light curve (including multiple oscillation modes and a higher chance of becoming triangles in the equivalent graph). In summary, the average clustering coefficient as a measure to the lightcurve roughness in time domain, in a similar way as connectivities in \citep{PG15b}, might be able to explain the higher injected energy to the HADS pulsations.
    
    \item The nodes degree distribution of HVG networks (Figure \ref{fig5} and \ref{fig6}) for \dsct\ stars obey the exponential distributions with indices higher than $\lambda=\ln(3/2)$ indicating the presence of correlated stochastic process in \dsct\ light curves. The lognormal behaviour for the light curve of \dsct\ stars networks also suggests a mechanism of a multiplicative generative process.  
    
     \item\ We find that the Kolmogorov–Smirnov distance ($KS_D$) between the HVG nodes degree distribution of a \dsct\ light curve and the analytical expression for an uncorrelated random model (with arbitrary random generation) can be an essential measure for cleaning frequencies to obtain the noisy residual of pulsating stars. The tiny $KS_D$ for a residual light curve after cleaning the frequencies may be a valuable criterion to validate the frequency analysis of pulsating stars (please see Figure \ref{fig9}).  
\end{itemize}

In conclusion, even if in our study we have several evidences of the exponential distribution form for PDF, that could be used to characterise the stochastic/chaotic behaviour of the light curves and also to study the corresponding pre-whitening cascade, our time series are short, and the populations for $k$ degree statistics have poor sampling with respect to other numerical studies \citep{lacasa2010, Kaki22,Adami24}. Further studies with longer time series are needed to strengthen the \textit{KS$_D$} distances as a pre-whitening criterion or to verify the presence of lognormal and/or power law distributions, and the underlying physical conditions where they emerge.

\begin{acknowledgements}
      This manuscript includes data collected by the \tess\ mission, which are publicly available from the Mikulski Archive for Space Telescopes (MAST). Funding for the \tess\ mission is provided by the NASA Explorer Program. We acknowledge financial support from project PID2019-107061GB-C63 from the ‘Programas Estatales de Generación de Conocimiento y Fortalecimiento Científico y Tecnológico del Sistema de I+D+i y de I+D+i Orientada a los Retos de la Sociedad’, from project PID2023-149439NB-C42 from the 'Proyectos de Generación de Conocimiento' and from the Severo Ochoa grant CEX2021-001131-S funded by MICIU/AEI/10.13039/501100011033 and FEDER, EU.
\end{acknowledgements}

\bibliographystyle{aa} 
\bibliography{bibtex.bib} 

\begin{thebibliography}{70}
\expandafter\ifx\csname natexlab\endcsname\relax\def\natexlab#1{#1}\fi

\bibitem[{{Acosta-Tripailao} {et~al.}(2021){Acosta-Tripailao}, {Past{\'e}n}, \&
  {Moya}}]{Acosta-Tripailao2021Entrp}
{Acosta-Tripailao}, B., {Past{\'e}n}, D., \& {Moya}, P.~S. 2021, Entropy, 23,
  470

\bibitem[{{Adami} {et~al.}(2024){Adami}, {Masoomy}, \&
  {Nattagh-Najafi}}]{Adami24}
{Adami}, V., {Masoomy}, H., \& {Nattagh-Najafi}, M. 2024, arXiv e-prints,
  arXiv:2412.12290

\bibitem[{{Aerts} {et~al.}(2010){Aerts}, {Christensen-Dalsgaard}, \&
  {Kurtz}}]{aerts2010asteroseismology}
{Aerts}, C., {Christensen-Dalsgaard}, J., \& {Kurtz}, D.~W. 2010,
  {Asteroseismology} (Springer Netherlands)

\bibitem[{{Alipour} {et~al.}(2022){Alipour}, {Safari}, {Verbeeck}, {Berghmans},
  {Auch{\`e}re}, {Chitta}, {Antolin}, {Barczynski}, {Buchlin}, {Aznar
  Cuadrado}, {Dolla}, {Georgoulis}, {Gissot}, {Harra}, {Katsiyannis}, {Long},
  {Mandal}, {Parenti}, {Podladchikova}, {Petrova}, {Soubri{\'e}},
  {Sch{\"u}hle}, {Schwanitz}, {Teriaca}, {West}, \& {Zhukov}}]{Alipour2022A&A}
{Alipour}, N., {Safari}, H., {Verbeeck}, C., {et~al.} 2022, \aap, 663, A128

\bibitem[{{Antoci}(2014)}]{Antoci2014}
{Antoci}, V. 2014, in IAU Symposium, Vol. 301, Precision Asteroseismology, ed.
  J.~A. {Guzik}, W.~J. {Chaplin}, G.~{Handler}, \& A.~{Pigulski}, 333--340

\bibitem[{{Antoci} {et~al.}(2011){Antoci}, {Handler}, {Campante}, {Thygesen},
  {Moya}, {Kallinger}, {Stello}, {Grigahc{\`e}ne}, {Kjeldsen}, {Bedding},
  {L{\"u}ftinger}, {Christensen-Dalsgaard}, {Catanzaro}, {Frasca}, {De Cat},
  {Uytterhoeven}, {Bruntt}, {Houdek}, {Kurtz}, {Lenz}, {Kaiser}, {van Cleve},
  {Allen}, \& {Clarke}}]{Antoci2011}
{Antoci}, V., {Handler}, G., {Campante}, T.~L., {et~al.} 2011, \nat, 477, 570

\bibitem[{{Baglin} {et~al.}(2009){Baglin}, {Auvergne}, {Barge}, {Deleuil},
  {Michel}, \& {CoRoT Exoplanet Science Team}}]{Baglin2009}
{Baglin}, A., {Auvergne}, M., {Barge}, P., {et~al.} 2009, in {IAU Symposium},
  Vol. {253}, {Transiting Planets}, ed. {{Pont}, Fr{\'e}d{\'e}ric and
  {Sasselov}, Dimitar and {Holman}, Matthew J.}, {71--81}

\bibitem[{{Baiesi} \& {Paczuski}(2004)}]{Baiesi-PRE}
{Baiesi}, M. \& {Paczuski}, M. 2004, \pre, 69, 066106

\bibitem[{Barabasi \& Oltvai(2004)}]{barabasi2004network}
Barabasi, A.-L. \& Oltvai, Z.~N. 2004, Nature reviews genetics, 5, 101

\bibitem[{Barabási \& Pósfai(2016)}]{barabasi2016}
Barabási, A.-L. \& Pósfai, M. 2016, Network science (Cambridge: Cambridge
  University Press)

\bibitem[{{Barcel{\'o} Forteza} {et~al.}(2015){Barcel{\'o} Forteza}, {Michel},
  {Roca Cort{\'e}s}, \& {Garc{\'\i}a}}]{Barcelo_Forteza2015}
{Barcel{\'o} Forteza}, S., {Michel}, E., {Roca Cort{\'e}s}, T., \&
  {Garc{\'\i}a}, R.~A. 2015, \aap, 579, A133

\bibitem[{{Bazarghan} {et~al.}(2008){Bazarghan}, {Safari}, {Innes}, {Karami},
  \& {Solanki}}]{Bazarghan2008A&A}
{Bazarghan}, M., {Safari}, H., {Innes}, D.~E., {Karami}, E., \& {Solanki},
  S.~K. 2008, \aap, 492, L13

\bibitem[{{Bedding} {et~al.}(2020){Bedding}, {Murphy}, {Hey}, {Huber}, {Li},
  {Smalley}, {Stello}, {White}, {Ball}, {Chaplin}, {Colman}, {Fuller},
  {Gaidos}, {Harbeck}, {Hermes}, {Holdsworth}, {Li}, {Li}, {Mann}, {Reese},
  {Sekaran}, {Yu}, {Antoci}, {Bergmann}, {Brown}, {Howard}, {Ireland},
  {Isaacson}, {Jenkins}, {Kjeldsen}, {McCully}, {Rabus}, {Rains}, {Ricker},
  {Tinney}, \& {Vanderspek}}]{Tim2020}
{Bedding}, T.~R., {Murphy}, S.~J., {Hey}, D.~R., {et~al.} 2020, \nat, 581, 147

\bibitem[{Boccaletti {et~al.}(2006)Boccaletti, Latora, Moreno, Chavez, \&
  Hwang}]{boccaletti2006}
Boccaletti, S., Latora, V., Moreno, Y., Chavez, M., \& Hwang, D.-U. 2006,
  Physics Reports, 424, 175

\bibitem[{{Bowman} \& {Kurtz}(2014)}]{bowman2014}
{Bowman}, D.~M. \& {Kurtz}, D.~W. 2014, \mnras, 444, 1909

\bibitem[{Braga {et~al.}(2016)Braga, Alves, Costa, Ribeiro, de Jesus,
  Tateishi, \& Ribeiro}]{BRAGA2016}
Braga, A., Alves, L., Costa, L., {et~al.} 2016, Physica A: Statistical
  Mechanics and its Applications, 444, 1003

\bibitem[{{Breger}(2000{\natexlab{a}})}]{Breger2000Balt}
{Breger}, M. 2000{\natexlab{a}}, Baltic Astronomy, 9, 149

\bibitem[{{Breger}(2000{\natexlab{b}})}]{Breger2000}
{Breger}, M. 2000{\natexlab{b}}, in ASP Conf. Series, Vol. 210, ASP Conf.
  Series, Vol.~210, 3

\bibitem[{{Breger}(2007)}]{Breger2007}
{Breger}, M. 2007, Communications in Asteroseismology, 150, 25

\bibitem[{Chang {et~al.}(2013)Chang, Protopapas, Kim, \& Byun}]{Chang2013}
Chang, S.-W., Protopapas, P., Kim, D.-W., \& Byun, Y.-I. 2013, Astron. J., 145,
  132

\bibitem[{{Christy}(1964)}]{Christy1964}
{Christy}, R.~F. 1964, Reviews of Modern Physics, 36, 555

\bibitem[{{Christy}(1966)}]{Christy1966ApJ}
{Christy}, R.~F. 1966, \apj, 144, 108

\bibitem[{{Daei} {et~al.}(2017){Daei}, {Safari}, \& {Dadashi}}]{Daei2017ApJ}
{Daei}, F., {Safari}, H., \& {Dadashi}, N. 2017, \apj, 845, 36

\bibitem[{{de Franciscis} {et~al.}(2018){de Franciscis}, {Pascual-Granado},
  {Su{\'a}rez}, {Garc{\'\i}a Hern{\'a}ndez}, \&
  {Garrido}}]{deFranciscis2018MNRAS}
{de Franciscis}, S., {Pascual-Granado}, J., {Su{\'a}rez}, J.~C., {Garc{\'\i}a
  Hern{\'a}ndez}, A., \& {Garrido}, R. 2018, \mnras, 481, 4637

\bibitem[{{de Franciscis} {et~al.}(2019){de Franciscis}, {Pascual-Granado},
  {Su{\'a}rez}, {Garc{\'\i}a Hern{\'a}ndez}, {Garrido}, {Lares-Martiz}, \&
  {Rod{\'o}n}}]{deFranciscis2019MNRAS}
{de Franciscis}, S., {Pascual-Granado}, J., {Su{\'a}rez}, J.~C., {et~al.} 2019,
  \mnras, 487, 4457

\bibitem[{{Farhang} {et~al.}(2022){Farhang}, {Shahbazi}, \&
  {Safari}}]{Farhang2022ApJ}
{Farhang}, N., {Shahbazi}, F., \& {Safari}, H. 2022, \apj, 936, 87

\bibitem[{{Feinstein} {et~al.}(2019){Feinstein}, {Montet}, {Foreman-Mackey},
  {Bedell}, {Saunders}, {Bean}, {Christiansen}, {Hedges}, {Luger}, {Scolnic},
  \& {Cardoso}}]{Feinstein2019}
{Feinstein}, A.~D., {Montet}, B.~T., {Foreman-Mackey}, D., {et~al.} 2019,
  \pasp, 131, 094502

\bibitem[{{Gheibi} {et~al.}(2017){Gheibi}, {Safari}, \&
  {Javaherian}}]{Gheibi2017ApJ}
{Gheibi}, A., {Safari}, H., \& {Javaherian}, M. 2017, \apj, 847, 115

\bibitem[{{Ghimire} {et~al.}(2020){Ghimire}, {Jadidoleslam}, {Krajewski}, \&
  {Tsonis}}]{Ganesh2020}
{Ghimire}, G.~R., {Jadidoleslam}, N., {Krajewski}, W.~F., \& {Tsonis}, A.~A.
  2020, Frontiers in Water, 2, 17

\bibitem[{{Gilliland} {et~al.}(2010){Gilliland}, {Brown},
  {Christensen-Dalsgaard}, {Kjeldsen}, {Aerts}, {Appourchaux}, {Basu},
  {Bedding}, {Chaplin}, {Cunha}, {De Cat}, {De Ridder}, {Guzik}, {Handler},
  {Kawaler}, {Kiss}, {Kolenberg}, {Kurtz}, {Metcalfe}, {Monteiro}, {Szab{\'o}},
  {Arentoft}, {Balona}, {Debosscher}, {Elsworth}, {Quirion}, {Stello},
  {Su{\'a}rez}, {Borucki}, {Jenkins}, {Koch}, {Kondo}, {Latham}, {Rowe}, \&
  {Steffen}}]{Gilliland2010}
{Gilliland}, R.~L., {Brown}, T.~M., {Christensen-Dalsgaard}, J., {et~al.} 2010,
  \pasp, 122, 131

\bibitem[{{Hasanzadeh} {et~al.}(2021){Hasanzadeh}, {Safari}, \&
  {Ghasemi}}]{Hasanzadeh2021}
{Hasanzadeh}, A., {Safari}, H., \& {Ghasemi}, H. 2021, \mnras, 505, 1476

\bibitem[{{Houdek} {et~al.}(1999){Houdek}, {Balmforth},
  {Christensen-Dalsgaard}, \& {Gough}}]{Houdek1999}
{Houdek}, G., {Balmforth}, N.~J., {Christensen-Dalsgaard}, J., \& {Gough},
  D.~O. 1999, \aap, 351, 582

\bibitem[{Iacobello {et~al.}(2018)Iacobello, Scarsoglio, \&
  Ridolfi}]{IACOBELLO20181}
Iacobello, G., Scarsoglio, S., \& Ridolfi, L. 2018, Physics Letters A, 382, 1

\bibitem[{{Kaki} {et~al.}(2022){Kaki}, {Farhang}, \& {Safari}}]{Kaki22}
{Kaki}, B., {Farhang}, N., \& {Safari}, H. 2022, Scientific Reports, 12, 16835

\bibitem[{{Kurtz} {et~al.}(2015){Kurtz}, {Shibahashi}, {Murphy}, {Bedding}, \&
  {Bowman}}]{Kurtz2015MNRAS}
{Kurtz}, D.~W., {Shibahashi}, H., {Murphy}, S.~J., {Bedding}, T.~R., \&
  {Bowman}, D.~M. 2015, \mnras, 450, 3015

\bibitem[{Lacasa {et~al.}(2008)Lacasa, Luque, Ballesteros, Luque, \&
  Nuno}]{lacasa2008}
Lacasa, L., Luque, B., Ballesteros, F., Luque, J., \& Nuno, J.~C. 2008,
  Proceedings of the National Academy of Sciences, 105, 4972

\bibitem[{Lacasa \& Toral(2010)}]{lacasa2010}
Lacasa, L. \& Toral, R. 2010, Phys. Rev. E, 82, 036120

\bibitem[{{Lares-Martiz}(2022)}]{Mariel2022}
{Lares-Martiz}, M. 2022, Frontiers in Astronomy and Space Sciences, 9, 301

\bibitem[{{Latora} \& {Marchiori}(2001)}]{Latora-PRL}
{Latora}, V. \& {Marchiori}, M. 2001, \prl, 87, 198701

\bibitem[{{Lightkurve Collaboration} {et~al.}(2018){Lightkurve Collaboration},
  {Cardoso}, {Hedges}, {Gully-Santiago}, {Saunders}, {Cody}, {Barclay}, {Hall},
  {Sagear}, {Turtelboom}, {Zhang}, {Tzanidakis}, {Mighell}, {Coughlin}, {Bell},
  {Berta-Thompson}, {Williams}, {Dotson}, \& {Barentsen}}]{Lightkurve}
{Lightkurve Collaboration}, {Cardoso}, J. V. d.~M., {Hedges}, C., {et~al.}
  2018, {Lightkurve: Kepler and TESS time series analysis in Python},
  Astrophysics Source Code Library, record ascl:1812.013

\bibitem[{{Lotfi} {et~al.}(2020){Lotfi}, {Javaherian}, {Kaki}, {Darooneh}, \&
  {Safari}}]{Lotfi2020Chaos}
{Lotfi}, N., {Javaherian}, M., {Kaki}, B., {Darooneh}, A.~H., \& {Safari}, H.
  2020, Chaos, 30, 043124

\bibitem[{Luque {et~al.}(2009)Luque, Lacasa, Ballesteros, \&
  Luque}]{luque2009horizontal}
Luque, B., Lacasa, L., Ballesteros, F., \& Luque, J. 2009, Physical Review E,
  80, 046103

\bibitem[{Mahmoud(2000)}]{mahmoud2000sorting}
Mahmoud, H. 2000, Sorting: A Distribution Theory, Wiley Series in Discrete
  Mathematics and Optimization (Wiley)

\bibitem[{Mathias \& Gopal(2001)}]{Mathias-PRE}
Mathias, N. \& Gopal, V. 2001, Phys. Rev. E, 63, 021117

\bibitem[{Mitzenmacher(2004)}]{mitzenmacher2004}
Mitzenmacher, M. 2004, Internet mathematics, 1, 226

\bibitem[{{Mohammadi} {et~al.}(2021){Mohammadi}, {Alipour}, {Safari}, \&
  {Zamani}}]{Mohammadi2021JGRA}
{Mohammadi}, Z., {Alipour}, N., {Safari}, H., \& {Zamani}, F. 2021, Journal of
  Geophysical Research (Space Physics), 126, e28868

\bibitem[{Mu{\~n}oz \& Garc{\'e}s(2021)}]{munoz2021}
Mu{\~n}oz, V. \& Garc{\'e}s, N.~E. 2021, Plos one, 16, e0259735

\bibitem[{{Nardetto} {et~al.}(2014){Nardetto}, {Poretti}, {Rainer}, {Guiglion},
  {Scardia}, {Schmid}, \& {Mathias}}]{Nardetto2014}
{Nardetto}, N., {Poretti}, E., {Rainer}, M., {et~al.} 2014, \aap, 561, A151

\bibitem[{Newman(2010)}]{newman2010}
Newman, M. 2010, Networks: An Introduction (OUP Oxford)

\bibitem[{Newman(2003)}]{newman2003}
Newman, M.~E. 2003, SIAM review, 45, 167

\bibitem[{{Pascual-Granado} {et~al.}(2015{\natexlab{a}}){Pascual-Granado},
  {Garrido}, \& {Su{\'a}rez}}]{PG15b}
{Pascual-Granado}, J., {Garrido}, R., \& {Su{\'a}rez}, J.~C.
  2015{\natexlab{a}}, \aap, 581, A89

\bibitem[{{Pascual-Granado} {et~al.}(2015{\natexlab{b}}){Pascual-Granado},
  {Garrido}, \& {Su{\'a}rez}}]{PG15a}
{Pascual-Granado}, J., {Garrido}, R., \& {Su{\'a}rez}, J.~C.
  2015{\natexlab{b}}, \aap, 575, A78

\bibitem[{{Pascual-Granado} {et~al.}(2018){Pascual-Granado}, {Su{\'a}rez},
  {Garrido}, {Moya}, {Garc{\'\i}a Hern{\'a}ndez}, {Rod{\'o}n}, \&
  {Lares-Martiz}}]{PG18}
{Pascual-Granado}, J., {Su{\'a}rez}, J.~C., {Garrido}, R., {et~al.} 2018, \aap,
  614, A40

\bibitem[{{Past{\'e}n} {et~al.}(2018){Past{\'e}n}, {Czechowski}, \&
  {Toledo}}]{Pasten2018Chaos}
{Past{\'e}n}, D., {Czechowski}, Z., \& {Toledo}, B. 2018, Chaos, 28, 083128

\bibitem[{{Pietronero} \& {Siebesma}(1986)}]{Pietronero1986}
{Pietronero}, L. \& {Siebesma}, A.~P. 1986, \prl, 57, 1098

\bibitem[{{Reegen}(2012)}]{Peter2012}
{Reegen}, P. 2012, Communications in Asteroseismology, 163, 3

\bibitem[{{Ricker} {et~al.}(2015){Ricker}, {Winn}, {Vanderspek}, {Latham},
  {Bakos}, {Bean}, {Berta-Thompson}, {Brown}, {Buchhave}, {Butler}, {Butler},
  {Chaplin}, {Charbonneau}, {Christensen-Dalsgaard}, {Clampin}, {Deming},
  {Doty}, {De Lee}, {Dressing}, {Dunham}, {Endl}, {Fressin}, {Ge}, {Henning},
  {Holman}, {Howard}, {Ida}, {Jenkins}, {Jernigan}, {Johnson}, {Kaltenegger},
  {Kawai}, {Kjeldsen}, {Laughlin}, {Levine}, {Lin}, {Lissauer}, {MacQueen},
  {Marcy}, {McCullough}, {Morton}, {Narita}, {Paegert}, {Palle}, {Pepe},
  {Pepper}, {Quirrenbach}, {Rinehart}, {Sasselov}, {Sato}, {Seager},
  {Sozzetti}, {Stassun}, {Sullivan}, {Szentgyorgyi}, {Torres}, {Udry}, \&
  {Villasenor}}]{Ricker2015}
{Ricker}, G.~R., {Winn}, J.~N., {Vanderspek}, R., {et~al.} 2015, Journal of
  Astronomical Telescopes, Instruments, and Systems, 1, 014003

\bibitem[{{Samadi} {et~al.}(2002){Samadi}, {Goupil}, \& {Houdek}}]{Samadi2002}
{Samadi}, R., {Goupil}, M.~J., \& {Houdek}, G. 2002, \aap, 395, 563

\bibitem[{{Souma} {et~al.}(2003){Souma}, {Fujiwara}, \&
  {Aoyama}}]{Souma2003PhyA}
{Souma}, W., {Fujiwara}, Y., \& {Aoyama}, H. 2003, Physica A Statistical
  Mechanics and its Applications, 324, 396

\bibitem[{{Steindl} {et~al.}(2021){Steindl}, {Zwintz}, \&
  {Bowman}}]{Steindl2021}
{Steindl}, T., {Zwintz}, K., \& {Bowman}, D.~M. 2021, \aap, 645, A119

\bibitem[{{Stellingwerf}(1980)}]{Stellingwerf1980}
{Stellingwerf}, R.~F. 1980, in Nonradial and Nonlinear Stellar Pulsation, ed.
  H.~A. {Hill} \& W.~A. {Dziembowski}, Vol. 125 (Springer Germany), 50--54

\bibitem[{{Su{\'a}rez} {et~al.}(2006){Su{\'a}rez}, {Garrido}, \&
  {Goupil}}]{Suarez06}
{Su{\'a}rez}, J.~C., {Garrido}, R., \& {Goupil}, M.~J. 2006, \aap, 447, 649

\bibitem[{{Su{\'a}rez} {et~al.}(2002){Su{\'a}rez}, {Michel}, {P{\'e}rez
  Hern{\'a}ndez}, {Lebreton}, {Li}, \& {Fox Machado}}]{Suarez02}
{Su{\'a}rez}, J.~C., {Michel}, E., {P{\'e}rez Hern{\'a}ndez}, F., {et~al.}
  2002, \aap, 390, 523

\bibitem[{{Sullivan} {et~al.}(2015){Sullivan}, {Winn}, {Berta-Thompson},
  {Charbonneau}, {Deming}, {Dressing}, {Latham}, {Levine}, {McCullough},
  {Morton}, {Ricker}, {Vanderspek}, \& {Woods}}]{Sullivan2015}
{Sullivan}, P.~W., {Winn}, J.~N., {Berta-Thompson}, Z.~K., {et~al.} 2015, \apj,
  809, 77

\bibitem[{{van Saders} \& {Pinsonneault}(2013)}]{Saders13}
{van Saders}, J.~L. \& {Pinsonneault}, M.~H. 2013, \apj, 776, 67

\bibitem[{Vanderspek {et~al.}(2018)Vanderspek, Doty, Fausnaugh,
  {et~al.}}]{Vanderspek2018}
Vanderspek, R., Doty, J.~P., Fausnaugh, M., {et~al.} 2018, TESS Instrument
  Handbook v0.1, technical Report

\bibitem[{Vogel {et~al.}(2020)Vogel, Brevis, Past{\'e}n, Mu{\~n}oz, Miranda, \&
  Chian}]{vogel2020measuring}
Vogel, E.~E., Brevis, F.~G., Past{\'e}n, D., {et~al.} 2020, Natural Hazards and
  Earth System Sciences, 20, 2943

\bibitem[{Watts \& Strogatz(1998)}]{Watts-nature1998}
Watts, D.~J. \& Strogatz, S.~H. 1998, Nature, 393, 440

\bibitem[{{Zou} {et~al.}(2019){Zou}, {Donner}, {Marwan}, {Donges}, \&
  {Kurths}}]{zou2019}
{Zou}, Y., {Donner}, R.~V., {Marwan}, N., {Donges}, J.~F., \& {Kurths}, J.
  2019, \physrep, 787, 1

\bibitem[{{Zou} {et~al.}(2014){Zou}, {Small}, {Liu}, \& {Kurths}}]{Zou14}
{Zou}, Y., {Small}, M., {Liu}, Z., \& {Kurths}, J. 2014, New Journal of
  Physics, 16, 013051

\end{thebibliography}

\begin{appendix}
\onecolumn
\section{HADS and LADS differences in time domain}\label{A}

\begin{figure*}[h!]
     \centering
     \includegraphics[width=8.5cm,height=6cm]
     {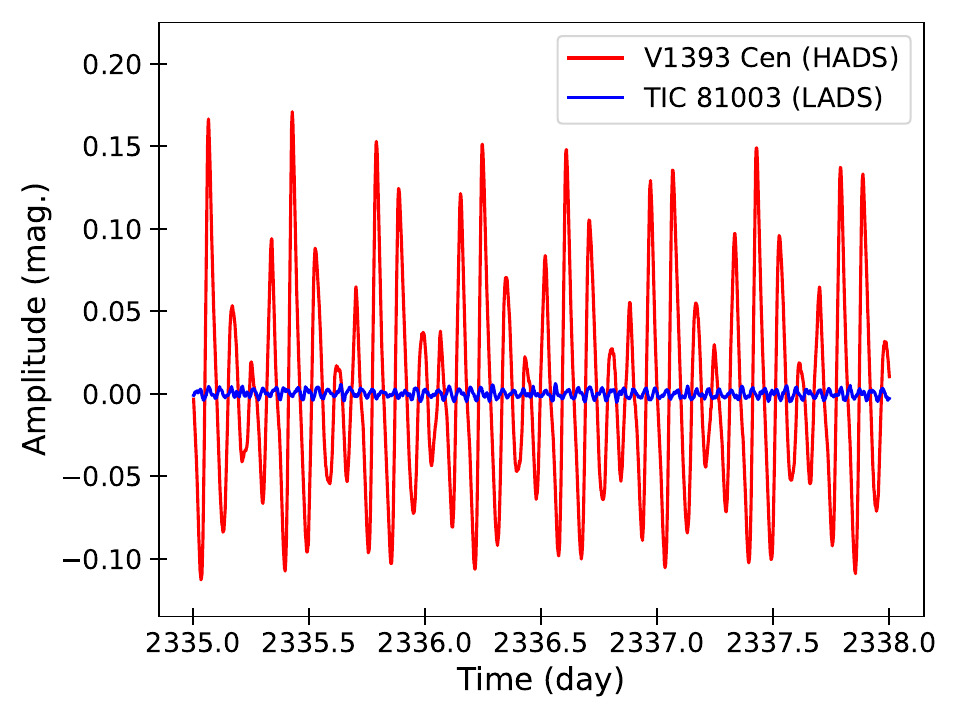}  
     \caption{Three-day light curves of a HADS (V1393 Cen) and a LADS (TIC 81003) observed by \tess, respectively in red and blue.}
     \label{figa1}
\end{figure*}

\section{Degree distribution differences in original, gap-filled, and cut light curves}\label{B}

\begin{figure*}[h!]
     \centering
     \includegraphics[width=8.5cm,height=6cm]
     {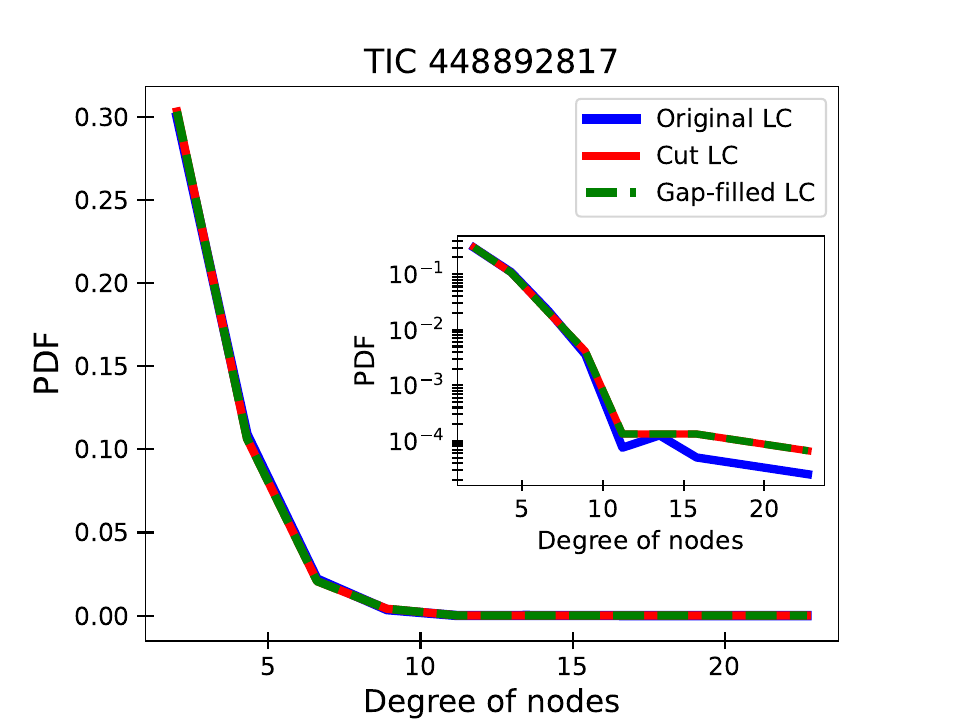}  
     \caption{PDF of HVG node degrees for original light curve (solid Blue), cut light curve (solid red), and gap-filled light curve (dashed green) of TIC 448892817. The distributions with logarithmic scale show the differences when the gaps exist.}
    \label{figb1}
\end{figure*}

\newpage

\begin{table}[htp]
\section{Degree distributions fitting parameters}\label{C}
\centering 
\caption{The list of HADS stars accompanied with the $\lambda$ indices of exponential fits and their uncertainties, p-value of exponential fits, $\mu$ and $\sigma$ of the lognormal fits and, the p-values of lognormal fits.}
\centering 
\begin{tabular}{ccccccccc}
\hline \hline
\multicolumn{9}{c} {Star ID} \\ 
\hline
 & Name & HD & TIC & $\lambda$ & p-value (exponential)& $\mu$ & $\sigma$ & p-value (lognormal)\\  
\cmidrule(l){1-9} 

1&AD Cmi &64191  &266328148 & -1.10$\pm$0.14 & 0.93& 1.35 & 0.25 & 1.0 \\
2&BE Lyn& 79889 &56914404 & -0.86$\pm$0.12 & 0.73& 1.34 & 0.028 & 0.86 \\
3&CY Aqr &--    &422412568 & -0.53$\pm$0.11 & 0.29& 1.33 & 0.32 & 0.42 \\
4&DX Cet &16189 &278962831 & -0.69$\pm$0.12 & 0.47& 1.35 & 0.24 & 0.34 \\ 
5&GW Uma &--    &150276417 & -0.69$\pm$0.07 & 0.99& 1.33 & 0.32 & 0.67 \\
6&PT Com &--   &335826251 & -0.92$\pm$0.06 & 1.0& 1.32 & 0.36 & 0.99 \\   
7&RS Gru&206379 &139845816 & -1.12$\pm$0.23 & 0.96& 1.33 & 0.30 & 0.99 \\
8&V 524 And &--  &196562983 & -1.0$\pm$0.15 & 0.98& 1.33 & 0.31 & 1.0 \\
9&V367 Cam &--   &354872568 & -0.89$\pm$0.05 & 0.98& 1.34 & 0.30 & 0.99 \\
10&V1162 Ori &-- &34512862 & -1.13$\pm$0.10 & 0.98& 1.34 & 0.29 & 0.99 \\
11&V2455 Cyg &204615 &266794067 & -0.60$\pm$0.14 & 0.16& 1.35 & 0.26 & 0.58 \\
12&AN Lyn &--   &56882581 & -0.86$\pm$0.11 & 0.86& 1.33 & 0.32 & 0.67 \\
13&--&146953& 210548440 & -0.94$\pm$0.07 & 0.98& 1.33 & 0.32 & 1.0 \\
14&V1384 Tau&--&415333069 & -0.88$\pm$0.10 & 0.98& 1.32 & 0.36 & 0.86 \\
15&V1393 Cen&121517&241787384 & -0.87$\pm$0.03 & 0.96& 1.34 & 0.28 & 1.0 \\
16&V2855 Ori&254061 &166979292 & -1.05$\pm$0.04 & 0.96& 1.34 & 0.29 & 1.0 \\
17&ZZ Mic&199757&126659093 & -1.47$\pm$0.41 & 0.96& 1.36 & 0.23 & 1.0 \\
18&AE Uma&--&357132618 & -0.73$\pm$0.03 & 0.99& 1.33 & 0.32 & 0.99 \\
19&$\rho$~ Pup&67523&154360594 & -0.93$\pm$0.19 & 0.47& 1.36 & 0.19 & 0.98 \\
20&
BL Cam&--    &392774261 & -0.72$\pm$0.05 & 0.99& 1.33 & 0.34 & 1.0 \\
21&BS Aqr&223338&9632550 & -0.76$\pm$0.23 & 0.98& 1.33 & 0.31 & 1.0 \\
22&DE Lac&--    &119486942 & -0.84$\pm$0.07 & 0.99 & 1.33 & 0.36 & 1.0  \\
23&KZ Hya&94033 &188209486 & -0.73$\pm$0.07 & 0.83& 1.33 & 0.31 & 0.92 \\
24&RY Lep&38882 &93441696 & -0.95$\pm$0.13 & 0.66& 1.35 & 0.24 & 0.99   \\
25&SS Psc&--&456857185 & -0.97$\pm$0.05 & 1.0 & 1.31 & 0.37 & 0.99  \\
26&SX Phe&223065&224285325 & -0.62$\pm$0.05 & 0.89 & 1.32 & 0.33 & 0.71 \\
27&V1719 Cyg&200925&290277380 & -1.11$\pm$0.05 & 0.96 & 1.35 & 0.26 & 1.0  \\
28&VX Hya&--&289711518 & -0.63$\pm$0.11 & 0.99 & 1.31 & 0.37 & 1.0 \\
29&VZ Cnc&73857&366632312 & -0.41$\pm$0.05 & 0.63 & 1.32 & 0.32 & 0.11 \\
30&XX Cyg&--    &233310793 & -0.71$\pm$0.08 & 0.99 & 1.31 & 0.38 & 1.0  \\
31&--&--& 448892817 & -0.80$\pm$0.16 & 0.86 & 1.32 & 0.36 & 0.93\\

\hline \hline
\end{tabular}
\label{tab1}
\end{table}

\begin{table*}[!h]
\begin{center}
\caption{The list of LADS stars accompanied with the $\lambda$ indices of exponential fits and their uncertainties, p-value of exponential fits, $\mu$ and $\sigma$ of the lognormal fits and, the p-values of lognormal fits.}
\centering 
\begin{tabular}{ccccccccc}
\hline \hline
\multicolumn{9}{c} {Star ID}   \\  
\hline
&Name & HD & TIC & $\lambda$ & p-value (exponential) & $\mu$ & $\sigma$ & p-value (lognormal)\\  
\cmidrule(l){1-9}  
  
1&--& 129831& 81003 & -0.69$\pm$0.04 & 0.99& 1.29 & 0.42 & 0.99 \\
2&--&77914& 975071 & -0.86$\pm$0.06 & 0.99 & 1.30 & 0.40 & 0.99 \\
3&--&112063& 9591460  & -0.75$\pm$0.03 & 0.98 & 1.32 & 0.37 & 1.0\\
4&RX Cae&28837& 7808834 & -1.0$\pm$0.21 & 0.9  & 1.33 & 0.29 & 0.99\\ 
5&--&21295& 12524129 & -0.50$\pm$0.04 & 0.99 & 1.26 & 0.49 & 1.0\\
6&--&79111& 18658256 & -0.58$\pm$0.02 & 0.99& 1.27 & 0.47 & 0.99 \\
7&--&86312& 26957587 &  -0.69$\pm$0.08 & 0.99 & 1.29 & 0.43 & 0.99 \\
8&--&181280& 30624832 & -0.55$\pm$0.03 & 0.99 & 1.27 & 0.48 & 1.0 \\
9&--&44930& 34737955  & -0.62$\pm$0.05 & 0.99 & 1.28 & 0.46 & 0.93\\
10&--&185729& 79659787  & -0.78$\pm$0.07 & 0.99 & 1.30 & 0.40 & 0.99\\
11&--&25674& 34197596  & -0.62$\pm$0.02 & 0.99 & 1.31 & 0.38 & 1.0\\
12&--&113221& 102192161  & -0.47$\pm$0.02 & 0.99 & 1.26 & 0.49 & 0.99\\
13&--&180349& 121729614  & -0.48$\pm$0.02 & 0.99 & 1.25 & 0.50 & 0.99\\
14&--&216728& 137796620  & -0.51$\pm$0.03 & 0.99 & 1.26 & 0.49 & 0.99\\
15&--&113211& 253296458  & -0.50$\pm$0.02 & 0.99 & 1.26 & 0.49 & 0.99\\
16&--&31322& 246902545  & -0.73$\pm$0.11 & 0.99 & 1.32 & 0.36 & 1.0\\
17&--&32433& 348792358  & -0.68$\pm$0.04 & 0.99 & 1.29 & 0.44 & 0.89\\
18&--&56843& 387235455  & -0.71$\pm$0.02 & 1.0 & 1.29 & 0.42 & 0.99\\
19&--&38597& 100531058  & -0.69$\pm$0.07 & 0.99 & 1.32 & 0.35 & 0.99\\
20&--&99302& 458689740  & -0.65$\pm$0.04 & 1.0 & 1.28 & 0.45 & 0.89\\
21&V479 Tau&24550& 459908110  & -0.61$\pm$0.02 & 0.99 & 1.28 & 0.46 & 0.99\\
22&V353 Vel&93298& 106886169 & -0.62$\pm$0.03 & 0.99 & 1.27 & 0.46 & 0.99 \\
23&--&17341& 122615966  & -0.68$\pm$0.06 & 0.99 & 1.29 & 0.42 & 0.99\\
24&--&183281& 137341551 & -0.58$\pm$0.02 & 0.99 & 1.26 & 0.49 & 0.99 \\
25&--&182895& 159647185  & -0.59$\pm$0.04 & 1.0 & 1.27 & 0.48 & 0.89\\
26&--&46722& 172193026  & -0.92$\pm$0.05 & 0.98 & 1.32 & 0.34 & 1.0\\
27&--&8043& 196921106  & -0.51$\pm$0.04 & 0.92 & 1.26 & 0.48 & 0.99\\
28&CC Gru&214441& 161172103  & -0.72$\pm$0.04 & 0.83 & 1.33 & 0.32 & 0.92\\
29&--&24572& 242944780  & -0.68$\pm$0.06 & 0.99 & 1.30 & 0.41 & 0.99\\
30&--&--& 274038922  & -0.83$\pm$0.06 & 0.98 & 1.31 & 0.39 & 0.99\\
31&IN Dra&191804& 269697721  & -0.80$\pm$0.03 & 0.99 & 1.30 & 0.40 & 0.89\\
32&GW Dra&--& 329153513  & -0.83$\pm$0.03 & 0.99 & 1.31 & 0.39 & 0.99\\
33&CP Oct&21190& 348772511  & -0.86$\pm$0.05 & 0.98 & 1.31 & 0.37 & 0.99\\
34&IO Dra&193138& 403114672  & -0.75$\pm$0.03 & 0.98 & 1.30 & 0.40 & 0.99\\
35&--&42005& 408906554  & -0.71$\pm$0.04 & 0.99 & 1.28 & 0.45 & 0.99\\
36&DE Cmi&67852& 452982723 & -0.89$\pm$0.20 & 0.99 & 1.32 & 0.36 & 0.99\\
37&V435 Car&44958&255548143  & -0.85$\pm$0.02 & 1.0 & 1.32 & 0.37 & 1.0 \\
38&V1790 Ori&290799&11361473  & -0.62$\pm$0.05 & 0.89 & 1.28 & 0.46 & 0.96\\
\hline \hline
\end{tabular}
\label{tab2}
\end{center}
\end{table*}

\end{appendix}
\end{document}